\documentclass[unnumsec,webpdf,contemporary,medium]{oup-authoring-template}%

\onecolumn 

\theoremstyle{thmstyleone}%
\theoremstyle{thmstyletwo}%
\theoremstyle{thmstylethree}%

\begin{document}

\journaltitle{Journal Title Here}
\DOI{DOI added during production}
\copyrightyear{YEAR}
\pubyear{YEAR}
\vol{XX}
\issue{x}
\access{Published: Date added during production}
\appnotes{Paper}

\firstpage{1}


\title[Survey Function on Scalar Regression]{Function on Scalar Regression with Complex Survey Designs}

\author[1$\ast$]{Lily Koffman\ORCID{0000-0003-1543-2896}}
\author[2]{Sunan Gao\ORCID{0000-0003-2345-1771}}
\author[3]{Xinkai Zhou\ORCID{0000-0001-8564-3491}}
\author[4]{Andrew Leroux\ORCID{0000-0002-6355-3580}}
\author[5]{Ciprian Crainiceanu\ORCID{0000-0001-6601-3881}}
\author[5]{John Muschelli III\ORCID{0000-0001-6469-1750}}

\address[1]{\orgdiv{Data Science}, \orgname{University of Virginia}, \orgaddress{\street{Charlottesville}, \postcode{22903}, \state{VA}, \country{USA}}}
\address[2]{\orgdiv{Epidemiology}, \orgname{Johns Hopkins Bloomerg School of Public Health}, \orgaddress{\street{Baltimore}, \postcode{21205}, \state{MD}, \country{USA}}}
\address[3]{\orgdiv{Statistics and Data Science}, \orgname{Beijing Normal - Hong Kong Baptist University}, \orgaddress{\state{Zhuhai}, \country{China}}}
\address[4]{\orgdiv{Biostatistics and Informatics}, \orgname{Colorado School of Public Health}, \orgaddress{\street{Aurora}, \postcode{80045}, \state{CO}, \country{USA}}}
\address[5]{\orgdiv{Biostatistics}, \orgname{Johns Hopkins Bloomerg School of Public Health}, \orgaddress{\street{Baltimore}, \postcode{21205}, \state{MD}, \country{USA}}}

\corresp[$\ast$]{Corresponding author. \href{email:koffmanlily@virginia.edu}{koffman@virginia.edu}}

\received{Date}{0}{Year}
\revised{Date}{0}{Year}
\accepted{Date}{0}{Year}



\abstract{Large health surveys increasingly collect high-dimensional functional data from wearable devices, and function on scalar regression (FoSR) is used to quantify the relationship between these functional outcomes and scalar covariates like age and sex. However, existing methods for FoSR fail to account for complex survey design. We introduce inferential methods for FoSR with complex survey designs. The approach combines fast univariate inference (FUI) developed for functional outcomes and survey sampling inferential methods developed for scalar outcomes. Our approach consists of three steps: (1) fit survey weighted GLMs at each point along the functional domain, (2) smooth coefficients along the functional domain, and (3) use balanced repeated replication (BRR) or Rao-Wu-Yue-Beaumont (RWYB) bootstrap to obtain pointwise and joint confidence bands for the functional coefficients. The approach is motivated by association studies between continuous physical activity data and covariates collected in the National Health and Nutrition Examination Survey (NHANES). A first-of-its-kind analytical simulation study and empirical simulation using NHANES data demonstrates that our approach performs better than existing methods that do not account for the survey structure. Finally, application of the approach in NHANES shows the practical implications of accounting for survey structure. The approach is implemented in the R package \texttt{svyfosr}. } 






\maketitle

\section{Introduction}
\label{s:intro}
Large epidemiological studies, such as the National Health and Nutrition Examination Survey (NHANES) \citep{NHANES} and the National Health and Aging Trends Study (NHATS) \citep{NHATS}, employ complex survey designs. These designs involve multistage stratified sampling because simple random sampling from a large geographic area is impractical and prohibitively expensive \citep{surveys_intro}. When survey design is properly accounted for in statistical analysis, conclusions from a survey can be generalized to the target population. However, failing to correctly account for the survey design can lead to estimation bias  and underestimation of variability. Accounting for survey design involves handling unequal sampling probabilities (via participant weights) and the correlation between individuals within the same geographic area \citep{nhanes_guidelines_2018}. 

Large studies increasingly collect high dimensional data, such as physical activity measurements from wearable devices. These measurements can be summarized and analyzed at various temporal resolutions: sub-second, second, minute, or day level. Functional data analysis techniques, including function on scalar regression (FoSR), are often used to quantify the relationship between physical activity profiles (functional responses) and scalar covariates such as age and sex \citep{fda_book}. Recently, \citep{cui_fui} proposed Fast Univariate Inference (FUI), a fast and computationally efficient method for FoSR. However, FUI and other existing methods for FoSR (e.g., penalized splines \citep{FAMM}) do not allow for incorporation of complex survey design information. Previous work on integrating survey design with functional data analysis has focused on estimation of mean or median curves \citep{means2, means3, medians} and scalar on function regression \citep{parker_bayesian_2023,survey_sofr}. As with standard (non-functional) models, failing to account for survey structure with functional data can lead to biased estimates,  underestimated uncertainty, and non-generalizable results.

Substantial literature and software exists for estimation and inference under complex survey designs in regression with scalar outcomes \citep{lumley_software, Lumley2017, survey_reg}. In this paper we extend the foundational inferential principles of survey sampling for scalar outcomes to functional outcomes. While the problem is complex, its solution consists of relatively simple, familiar inferential steps. First, we fit a separate generalized linear model (GLM) at each point along the functional domain. Therefore, the theoretical and practical properties of the estimators are preserved at every point in the functional domain, including marginal consistency of point estimators and nominal coverage of pointwise confidence intervals.
Second, we use smoothing along the functional domain \citep{Crainiceanu_2012,fda_book,cui_fui,Park_Staicu_2018} to obtain the smooth estimators. Third, we use survey-aware replication methods for inference: balanced repeated replication (BRR) \citep{BRR} and the Rao-Wu-Yue-Beaumont (RWYB) bootstrap \citep{RWYB}; see our \texttt{R} \citep{R} package \texttt{svyfosr} for implementation of these methods. This approach - relying solely on marginal regressions for estimation and resampling techniques for inference - allows for the construction of pointwise and correlation and multiplicity adjusted (CMA) confidence intervals \citep{fda_book} for the functional estimators. The combination of these ideas leads to a highly computationally efficient algorithm. To validate our approach, we use two simulations. In the first, we design a novel analytical simulation to mimic two-stage stratified sampling with both correlated functional outcomes and informative sampling. In the second, we design an empirical simulation by informatively sampling functional outcomes from NHANES. Finally, we illustrate the practical implications of accounting for survey structure with an analysis of physical activity data in NHANES 2011-2012 and 2013-2014 waves. 
 
\section{Methods}\label{sec:meth}
\subsection{Review of FUI}\label{subsec:fui}

FUI was proposed for function on scalar regression with longitudinal functional data \citep{mfpca,lfpca,ivanescu_2015,Morris_Carroll_2006,Morris_2008,FAMM,sfpca}. Due to the complexity of survey weighting in the longitudinal setting \citep{survey_mixed}, we focus on survey methods for function on scalar regression with only one function per participant in this paper. In Web Appendix A, we include a simulation demonstrating the properties of FUI in the single level setting. Here, we provide a short summary of the approach and data structure. Let $Y_{i}(s)$ denote the functional outcome for study participant $i$, $i= 1, \dots, I$ at location $s$ in the functional domain $s \in \{s_1, s_2, \dots, s_L\}$. In the NHANES physical activity data, $Y_{i}(s)$ denotes the mean physical activity for participant $i$ at minute $s$, and $L=1440$ for all participants. We will assume that $Y_{i}(s)\sim{\rm EF}\{\eta_{i}(s_l)\}$, where ${\rm EF}(\cdot)$ denotes an exponential family of distributions, and $\eta_{i}(s_l)$ is its linear predictor. Let $\mathbf{X}_{i}$ be a $P$-dimensional vector of fixed effects. The single level FoSR model is
\begin{equation}
h\{E[Y_{i}(s_l)]\}=\mathbf{X}_{i}\boldsymbol{\beta}(s_l)\;,
\label{eq:FUI_sl}
\end{equation}
where $h(\cdot)$ is the link function and $\boldsymbol{\beta}(s_l)$ are the domain-varying fixed effects of the covariates.

FUI starts with fitting $L$ separate GLMs, one for each location $s_l$ in the functional domain. Fitting these models does not make any assumption about correlation across $s_l$. The second step of FUI is to smooth the estimated fixed effects, $\boldsymbol{\beta}(\cdot)$, along the functional domain under the independence of residuals assumption. The third step is to use a bootstrap sample of study participants to construct pointwise and joint confidence intervals for the smoothed estimators of $\boldsymbol{\beta}(\cdot)$.  The method works because: (1) the point estimators are consistent under weak standard assumptions; and (2) the bootstrap of subjects accounts for the variability of the estimators, the procedure, and the within-person correlation.

The goal of this paper is to extend FUI to the case when data are collected using a known survey sampling procedure instead of an assumed independent sample. There are two crucial components impacted by this change that need to be addressed for a successful extension of the method. First, a survey-weighted procedure must be used for the GLM portion of the algorithm. This implementation is relatively easy to do because there is stable software for survey-weighted GLMs. Second, the subject-specific bootstrap approach needs to be adjusted to account for the actual sampling procedure from the target population. We now provide the details of this combined approach.

\subsection{Function on scalar regression for survey data}\label{subsec:fosr}
We introduce notation for a two-stage survey design where individuals belong to both a stratum (larger geographic location) and primary sampling unit (PSU; a smaller area within the stratum, such as a zip code). PSUs are sometimes called clusters; in this paper we use PSU. Within each PSU, participants may be sampled either at random or with probabilities that are a function of certain characteristics. We discuss this survey design, as it reflects our data example, but methods can be extended to other, more complex designs.

Assume that the study contains $H$ strata and $C_h$ PSUs within each stratum, $h=1,\ldots,H$. $C_h$ is typically the same for all strata, but the index $h$ indicates that this is not required.  Individual observations are indexed by $(h,c,i)$, that is stratum $h$, PSU $c$ and individual $i$. The total number of individuals within each stratum-PSU combination is $I_{h,c}$ and the total number of individuals in the sample is $I = \sum_{h=1}^H\sum_{c=1}^{C_h}I_{h,c}$.

The data for a single participant is the functional outcome $Y_{hci}(s)$, a $P$-dimensional vector of covariates $\mathbf{x}_{hci}$, and the scalar survey weight $w_{hci}$. The observed data are then: $(\mathbf{Y}, \mathbf{X}, \mathbf{w}) = \big\{(Y_{hci}(s), \mathbf{x}_{hci}, w_{hci}): h = 1, \dots, H; c = 1, \dots, C_h; i = 1, \dots, I; s \in \{s_1, \dots, s_L\}\big\}$. The survey weights are the inverse probabilities of selection from the population; they can be thought of as the number of people in the population represented by that study participant. The sum of all weights $\sum_{h=1}^H\sum_{c=1}^{C_h}\sum_{i=1}^{I_{c,h}}w_{hci}=N$, the total population size in the target population (e.g., total US population). Recall that $I<<N$, as $w_{hci}>>1$ for all $h$, $c$, and $i$.

\subsection{Estimation}\label{subsec:estimation}
The target of inference is $\beta_p(s)$, for $p=1,\ldots,P$. 
As described in Section~\ref{subsec:fui}, the FUI approach provides initial estimators $\tilde{\beta}_p(s)$ by fitting model~\eqref{eq:FUI_sl} at every $s$ using survey-weighted GLM \citep{Lumley2017,lumley_software, SAS2023, Stata2025}. The resulting estimators are then smoothed as a function of $s$ using any smoothing procedure \citep{fda_book,Crainiceanu_2012};  we use generalized additive models [GAMs] based on penalized splines \citep{wahba1990spline, Wood2017, Ruppert2003, eilers1996flexible, osullivan}. This procedure provides point estimators of the smoothed functional coefficients and confidence intervals under the assumption of independence of residuals. Because residuals are not independent, we only use the smooth point estimators and not the confidence intervals, which do not have the correct coverage probability.

\subsection{Inference}\label{subsec:inference}
For inference, we use the smooth estimators and a bootstrap of study participants to obtain pointwise and CMA confidence intervals. Because study participants are not sampled independently in our analysis, we will use resampling techniques inspired by the survey sampling literature \citep{BRR,RWYB, raowuyue1992} for non-functional analyses. Thus, we take advantage of decades of research to conduct resampling and assess the variability of functional estimators. We now provide details of the exact implementation of these resampling techniques. All  techniques refer to how to resample the $I$ study participants from the observed data. Standard bootstrap refers to independent, equal probability sample of the index set $1,\ldots,I$ with replacement. We now describe three well-established methods in the survey sampling literature.

\subsubsection{Inference: survey weighted bootstrap}\label{subsubsec:survey}
The survey-weighted bootstrap is defined as random sampling with replacement of the index set $1, \dots, I$ where each individual's probability of selection is $\pi_{hci} \propto w_{hci}$, where $w_{hci}$ is the individual's survey weight. This can be implemented directly in most analytic software.

\subsubsection{Inference: Balanced Repeated Replication}\label{subsubsec:BRR}
Balanced Repeated Replication (BRR) \citep{BRR} is a resampling method that requires an even number of PSUs per stratum (the structure used in NHANES and NHATS). The BRR procedure is as follows: (1) randomly sample half of the PSUs from each stratum; (2) form the data set by combining all observations from all participants in the selected PSUs; (3) for all selected participants, double the survey weights; and (4) fit the model using these adjusted weights. The BRR procedure is then repeated $B$ times. The even PSUs per stratum requirement is reasonable in practice: many large health studies include two PSUs per stratum because it is the minimum required to estimate within-stratum variance. Applying BRR to a design with an uneven number of PSUs per stratum would lead to biased variance estimates. In this case, practitioners could either use a different method for variance estimation or modify the data to obtain the appropriate number of PSUs per stratum either by (a) collapsing similar strata together or (b) splitting single PSUs \citep{Korn1999}.

\subsubsection{Inference: Rao-Wu-Yue-Beaumont Bootstrap}\label{subsubsec:RWYB}
The Rao-Wu-Yue-Beaumont (RWYB) \citep{RWYB} is a bootstrap method for any multistage sampling design provided that selection probabilities are available at each stage of sampling. It extends the Rao-Wu-Yue method \citep{raowuyue1992} to account for settings where PSUs are sampled without replacement within strata. Here we use the method as described by the authors \citep{RWYB} and implemented in the {\ttfamily R} \citep{R} function {\ttfamily svrep::make\_rwyb\_bootstrap\_weights} \citep{svrep}. Though RWYB is referred to as a bootstrap approach, the method resembles reweighting: the result is $B$ sets of weights for the dataset. Survey-weighted regression is run with each set of weights and results are then aggregated across the $B$ sets to conduct inference. For completeness, we provide the main steps of the procedure and refer to the original paper \citep{RWYB} for additional technical details. The algorithm steps are: (1) generate first stage bootstrap adjustments for all PSUs; (2) generate the second-stage adjustments to account for the probability of selecting the higher-level unit in the previous stage; and (3) obtain the overall bootstrap weights by multiplying the original weight by the first and second stage adjustments. The approach can be used  for different sampling schemes, but here we focus on sampling without replacement with unequal probabilities at the first stage and Poisson sampling at the second stage, an approximation of the NHANES structure. 

\noindent The stage one bootstrap adjustments are: 
$$a_{1c} = 1 - \sqrt{\frac{m_1(1-\pi_{1c})}{n_1-1}} +\sqrt{\frac{m_1(1-\pi_{1c})}{n_1-1}} \frac{n_1}{m_1}m^*_{1c} \;,$$
where $n_1$ is the number of PSUs sampled without replacement in the stratum (determined by the survey design), $m_1$ is the number of bootstrap draws used to reselect the PSUs, and  $\pi_{1c}$ is the probability of selecting PSU $c$. The probability of selecting PSU $c$ depends on the type of sampling used. Here $m^*_{1c}$ is a random variable  equal to the number of times PSU $c$ is chosen in $m_1$ independent trials, where each $c$ has a constant probability of being chosen at each draw equal to $\frac{1}{n_1}$ (i.e. $m^*_{1c}\sim\mathrm{Multinom}(m_1, 1/n_1)$).
Each calculation is nested within a stratum and thus all parameters are indexed by $h$, but for simplicity we focus on just one stratum. The authors further propose to calibrate the first stage bootstrap weights to 
$a^{\rm cal}_{1c} = a_{1c}\frac{n_1}{\sum_{l}a_{1l}}$,
which ensures that  $n_1=\sum_{c}a^{\rm cal}_{1c}$.
The second-stage preliminary bootstrap adjustments are $$a_{2ci} = 1-\sqrt{\frac{\pi_{1c}}{2-\pi_{1c}}} + \sqrt{\frac{\pi_{1c}}{2-\pi_{1c}}}\tilde{a}_{2ci}\;,$$ 
where $\tilde{a}_{2ci}$ are random samples from a Gamma distribution with shape parameter $\frac{1}{1-\pi_{2ci}}$ and scale parameter $1-\pi_{2ci}$, and $\pi_{2ci}$ is the probability of selecting unit $i$ in PSU $c$. 
 The final bootstrap weights are $w^*_{12ci} = w_{12ci}a_{1c}a_{2ci}$.

\subsubsection{Inference: Confidence bands}\label{subsubsec:CMA} All resampling techniques produce samples of the smoothed estimators $\widehat{\beta}_p(s)$, which can be used to estimate the marginal variance ${\rm Var}\{\widehat{\beta}_p(s)\}$ at every $s$ and for every $p$.
Pointwise $1-\alpha$ level confidence intervals for the functional coefficient can then be obtained using $\widehat{\beta}_p(s)\pm z_{1-\alpha/2}\sqrt{{\rm Var}\{\widehat{\beta}_p(s)\}}$.  CMA confidence intervals can be obtained using one of the methods described in \citep{fda_book}. Here we provide the summary of an approach based on the assumption of joint normality of the sample $\widehat{\beta}_p^b(s)$ over all $s\in S$, where $b=1,\ldots,B$ indicates the bootstrap sample. 
The  procedure involves (1) using the $B$ bootstrap replicates of $\beta(s)$ to estimate the $L\times L$ correlation matrix $\mathbf{C}$, then (2) simulating from the multivariate normal distribution $N(0, \mathbf{C})$, and (3) calculating the max absolute statistic from each simulated draw, and (4) obtaining the 95th percentile of these max absolute statistics. Thus, the $95$\% CMA confidence intervals are given by $\widehat{\beta}_p(s)\pm q_{0.95}\sqrt{{\rm Var}\{\widehat{\beta}_p(s)\}}$, where $q_{0.95}$ is the $0.95$ quantile of the max absolute statistic of a vector with distribution $\mathcal{N}(0,\mathbf{C})$. See Chapter 2.4.2 in \citep{fda_book} and Section 4 in \citep{cui_fui} for more details. 
\subsection{Computational considerations}\label{subsec:comp_cons}
The algorithm for fitting survey FoSR is summarized below.

\begin{algorithm}[!ht]
\caption{Survey-weighted Functional-on-Scalar Regression (Survey FoSR)}
\label{alg:survey_fosr}
\vspace{0.1em}
\textbf{Estimation:}
\begin{enumerate}
    \item Fit a survey-weighted generalized linear model (GLM) at each point 
    $s_l$, $l = 1, \dots, L$, along the functional domain:
    \[
    h\{E_w[Y_i(s_l)]\} = \mathbf{X}_i \boldsymbol{\beta}(s_l),
    \]
    and obtain the estimated coefficients $\tilde{\boldsymbol{\beta}}(s_l)$.
    \item Smooth $\tilde{\boldsymbol{\beta}}(s_l)$ across $s$ to obtain 
    $\hat{\boldsymbol{\beta}}(s)$.
\end{enumerate}

\vspace{0.3em}
\textbf{Inference:}
\begin{enumerate}
    \item[3.] For $b = 1, \dots, B$:
    \begin{enumerate}
        \item[(a)] If using BRR: obtain a BRR replicate sample and corresponding replicate weights
        \item[(b)] If using RWYB: obtain RWYB replicate weights.
        \item[(c)] If using weighted: obtain bootstrap sample by sampling with replacement with probability proportional to survey weight
        \item[(d)] Refit steps (1)--(2) using the $b$th replicate to obtain 
        $\hat{\boldsymbol{\beta}}(s_l)_{(b)}$ for $l = 1, \dots, L$.
    \end{enumerate}
    \item[4.] For each $l = 1, \dots, L$, estimate 
    $\mathrm{Var}\{{\boldsymbol{\beta}}(s_l)\}$ 
    using the sample variance of the $B$ bootstrap estimates.
\end{enumerate}
\end{algorithm}

Fitting a separate regression at each point along the functional domain $L$ and repeating this procedure many times for bootstrap inference can be computationally burdensome. Our implementation leverages batching (fitting $L$ models simultaneously) and the QR decomposition to achieve substantial computational speedups. We will briefly discuss these ideas. 

\subsubsection{Computational considerations: Batching}
Consider first the case of linear regression, and then we will show how to extend ideas to GLMs. The first observation is that for this analysis the regression design matrix, $X$, is the same at every point of the functional domain. If the design changes with the functional domain, this method cannot be batched in this way.  The second observation is that fitting $L$ separate models is equivalent to computing the matrix multiplication $\widehat{B} = (X'WX)^{-1}X'WY$, where $\mathbf{Y}$ is the $n \times L$ matrix obtained by binding the outcomes at $L$ separate points in the functional domain. Here
$X$ is the $n\times p$ dimensional matrix of covariates and $W$ is a $n \times n$ diagonal matrix of weights. As $(X'WX)^{-1}$ does not depend on the outcome, this operation is computed only once. 
\subsubsection{Computational considerations: QR decomposition}
Instead of calculating $(X'WX)^{-1}$ directly, we use the QR decomposition $X_w = W^{1/2}X = QR$ \citep{gentle1998qr}. Then the coefficients are given by: 
$\widehat{B} =R^{-1}Q'Y_w$ where $Y_w = W^{1/2}Y$.

These ideas can be generalized to the GLMs, which are fit using iteratively reweighted least squares (IRLS) or Newton-Raphson. We extend the batching and QR decomposition approach to the GLM case  at each iteration of IRLS or Newton-Raphson. Because the procedure is the same, all iterations can be conducted simultaneously by reusing the weighted predictor matrix $X_w$ across iterations. 

\subsubsection{Computational considerations: Run time}
For a Gaussian dataset with $n=100$, $p=10$, and $L = 100$, the standard approach of fitting each regression separately using \texttt{glm.fit} in \texttt{R} required an average of $175$ ms, whereas our batched implementation with QR decomposition required just $5$ ms: a 35-fold improvement in computational efficiency.
For a non-Gaussian (Bernoulli) dataset with $n=100$, $p=10$, and $L=100$, the standard approach required an average of $560$ ms, whereas the batched approach took $430$ ms, representing a $30$\% improvement in efficiency. 
\subsection{Software}
The software to implement our approach is provided in the R package \texttt{svyfosr} (\url{https://github.com/jhuwit/svyfosr}). The syntax mirrors that of the \texttt{fastFMM::fui()} function:

\begin{verbatim}
model_fit = svyfosr::svyfui(Y ~ X1 + X2, data = sample_df, 
                            weights = weight, family = gaussian(),
                            boot_type = "BRR",num_boots = 100,
                            parallel = TRUE, n_cores = 6, seed = 2213)
\end{verbatim}
Here \texttt{Y~X1 + X2} is a model formula where \texttt{Y} is a matrix of functional outcomes in \texttt{sample\_df} and \texttt{X1} and\texttt{X2} are a columns of covariates in \texttt{sample\_df}. The survey weights argument \texttt{weights} are a column in \texttt{sample\_df} or an external vector, family is a GLM family object (e.g. \texttt{gaussian()}, \texttt{poisson()}), \texttt{binomial()}), and \texttt{boot\_type} denotes the type of bootstrap: unweighted, weighted, BRR, or RWYB. The number of bootstraps samples is given by \texttt{num\_boots}, and the final options allow for parallelization of the bootstrapping step and a seed for reproducibility.

\subsection{Theoretical considerations}

Our approach is a combination of existing methods and inherits its properties from each of these methods. Below, we summarize the assumptions required for our method. Assumptions relate to survey design and the model.

Survey design: 
\begin{itemize}
  \item[D1)] Survey weights are bounded 
    \item[D2)] Inclusion probabilities are positive and correctly specified 
    \item[D3)] BRR: even PSUs per stratum; RWYB: inclusion probabilities at each stage are known 
\end{itemize}

Model: 
\begin{itemize}
    \item[M1)]  $\beta(s)$ is smooth
    \item[M2)] The mean is correctly specified: $E(Y_i(s) | X_i) = h^{-1}(X_i\beta(s))$
    \item[M3)] The functional outcomes $Y_i(s)$ are independent conditional on covariates and the survey design 
\end{itemize}

With these assumptions, the following properties are guaranteed:

\textit{Consistency}: At each point along the functional domain, the survey-weighted pointwise estimator $\tilde{\beta}(s_l)$ is consistent for the population parameter due to standard survey asymptotics \citep{Wooldridge1999}. Under assumption M1, the smoothed estimator $\tilde{\beta}(s_l)$ is consistent for $\beta(s_l)$ due to results for penalized splines \citep{Xiao2019}, provided that (i) the smoothing parameter $\lambda$ approaches zero at an appropriate rate and (ii) the number of basis functions $k$ is sufficient to approximate $\beta(s)$. Condition (i) is guaranteed because GCV is applied to $\tilde{\beta}(s_l)$ which has pointwise variance $O(n^{-1})$ so $\lambda$ goes to zero as $n$ increases. For condition (ii), the number of basis functions should be large enough to capture the complexity of $\beta(s)$: a default of $k = \min(L/2, 35)$ is typically adequate for smooth coefficient functions, although users may increase as needed. See Web Appendix E for examples of when this does not hold.

\textit{Asymptotic coverage}: Pointwise confidence intervals achieve the nominal asymptotic coverage rate due to the consistency of the BRR and RWYB variance estimators under D1-D3, provided that the bias of the smoother is smaller than the standard error. Joint CMA intervals achieve the nominal asymptotic coverage rate under the additional assumption of joint asymptotic normality of $\hat{\beta}(s)$ across the functional domain \citep{fda_book}

We defer a full asymptotic treatment with convergence rates and optimal smoothing parameter selection to later work.

A note on smoothing order. Our approach obtains pointwise coefficient estimates $\hat{\beta}$ and smooths the estimates to obtain $\tilde{\beta}$. An alternative approach would be to smooth the functional outcomes first, then fit pointwise models to the smoothed data. With Gaussian data and the same linear smoother, the approaches are algebraically equivalent. However, they differ in how the smoothing parameter is selected. When smoothing is applied to the estimated coefficients $\tilde{\beta}$, whose pointwise variance is $\sigma^2/n$, GCV selects a smaller smoothing parameter as $n$ grows, ensuring the estimator is consistent. When smoothing is applied first, GCV selects the smoothing parameter from the outcomes, whose pointwise variance is approximately $\sigma^2$. This induces $O(1)$ bias that does not change with sample size. We demonstrate this phenomenon in Web Appendix E.  
\section{Simulations}\label{sec:sim}
\subsection{Theoretical simulation}\label{subsec:theoretical}
Our simulation procedure is novel and complex. {It is the first to combine the generation of functional outcomes in a complex survey structure with informative sampling. Previous work has either generated functional outcomes without survey structure, assuming the data are independent, \citep{cui_fui}, or generated functional predictors and scalar outcomes and defined informative sampling in this context \citep{ parker_bayesian_2023,survey_sofr}.  Our approach addresses three novel challenges: (1) generation of a realistic ``superpopulation'' that exhibits stratum- and PSU-specific correlation in functional outcomes using B-spline hierarchical random effects, (2) definition and implementation of informative sampling with functional outcomes, and (3) stratum-specific heterogeneity in covariate effects.} Here we provide the intuition behind its structure; in Web Appendix B we provide the R code for the simulation, and in Web Appendix C provide a vignette that walks through the procedure step-by-step with code and plots.

The simulation consists of three steps: (1)  generation of a superpopulation, (2) sampling from that superpopulation, and (3) fitting models to the sample. In the superpopulation generation step, a population of individuals who belong to a stratum and PSU are generated. Each individual has a functional outcome and a covariate, and there is heterogeneity across strata and PSUs. No survey design information is used to generate the superpopulation. The functional outcomes are (optionally) correlated within strata and PSUs, a key aspect that distinguishes survey data from independent data. In the sampling step, we select individuals from the superpopulation in an (optionally) informative manner. Here, the survey design dictates how we obtain our sample from the superpopulation. We then fit models on the sampled dataset, and compare the model estimates to the true data generating mechanism in the superpopulation. For the models using RWYB and BRR for inference, the survey design informs how the replicate weights are generated.

\subsubsection{Theoretical simulation: generation of the superpopulation}\label{subsubsec}

We generate a superpopulation of size $N=10^7$, where each individual in the superpopulation belongs to a stratum and PSU. Each individual in the superpopulation is assigned to a stratum, which can be thought of as a geographical area, and a PSU within that stratum, which can be thought of as a county or city. Dirichlet distributions are used to generate the probabilities of assignment to each stratum and each PSU within the stratum. We set $H=30$ strata and for each strata the number of PSUs in the stratum is sampled uniformly between $75$ and $125$. Stratum assignment probabilities are generated from a symmetric Dirichlet distribution with concentration parameter $4$, and PSU assignment probabilities are generated from a symmetric Dirichlet distribution with concentration parameter $10$. 

Each individual is then assigned a covariate independently sampled from a Normal distribution: $X_i \sim \mathcal{N}(0,2)$ and a baseline linear predictor: 
$$\eta_i(s) = h\{\mu_i(s)\} = \beta_0(s) + X_i\beta_1(s)\;,$$ where $\beta_0(s)$ is a global intercept function and $\beta_1(s)$ is a slope function. In particular, $\beta_0(s) = -0.15 - 0.10\sin(2\pi s) - 0.1\cos(2\pi s)$ and $\beta_1(s) = \frac{1}{20}\phi\Big(\frac{s-0.6}{0.0225}\Big)$, where $\phi(\cdot)$ is the standard Normal PDF.

Next, we create stratum-specific heterogeneity in the relationship between $X_i$ and the outcome by scaling $\beta_1(s)$ within each stratum, and we induce within-stratum and within-PSU correlations by simulating random functional effects at both the stratum and PSU levels, and adding them to the baseline linear predictor. Both of these steps are optional and create correlation within strata and PSU that often occur in real world survey data. The linear predictor for individual $i$ after this step is: 
$$\eta_i(s) = \beta_0(s) + X_i\beta_{1(h)}(s) + b_{h(i)}(s) + b_{c(h, i)}(s)\;.$$
Where $\beta_{1(h)}(s)$ is the stratum-specific slope function, $b_{h(i)}(s)$ is the stratum-specific functional random effect, and $b_{c(h, i)}(s)$ is the PSU-specific functional random effect. 

The stratum-specific slope function is created by scaling the global slope function: $\beta_{1(h)}(s) = \gamma_h \cdot \beta_1(s)$, where $\gamma_h \sim \mathcal{N}(1, \sigma_s^2)$. The stratum and PSU-specific random effects are modeled as smooth functions generated from B-spline bases. The random effect for stratum $h$ is $b_h(s) = \sum_{k=1}^K\xi_{hk}\phi_k(s)$ and the PSU-level random effect for PSU $c$ nested within stratum $h$ is $b_{c(h)}(s) = \sum_{k=1}^K\zeta_{gk}\phi_k(s)$, where $\{\phi_k(s)\}_{k=1}^K$ is a B-spline basis of dimension $K=5$ over $[0,1]$. The $\xi_{hk}$ and $\zeta_{hk}$ are drawn from mean-zero Normal distributions with variances $\sigma^2_h$ and $\sigma^2_h/2$, respectively, such that the stratum-level random effects contribute more variation than the PSU-level random effects.

\begin{figure}
    \centering
    \includegraphics[width=\textwidth]{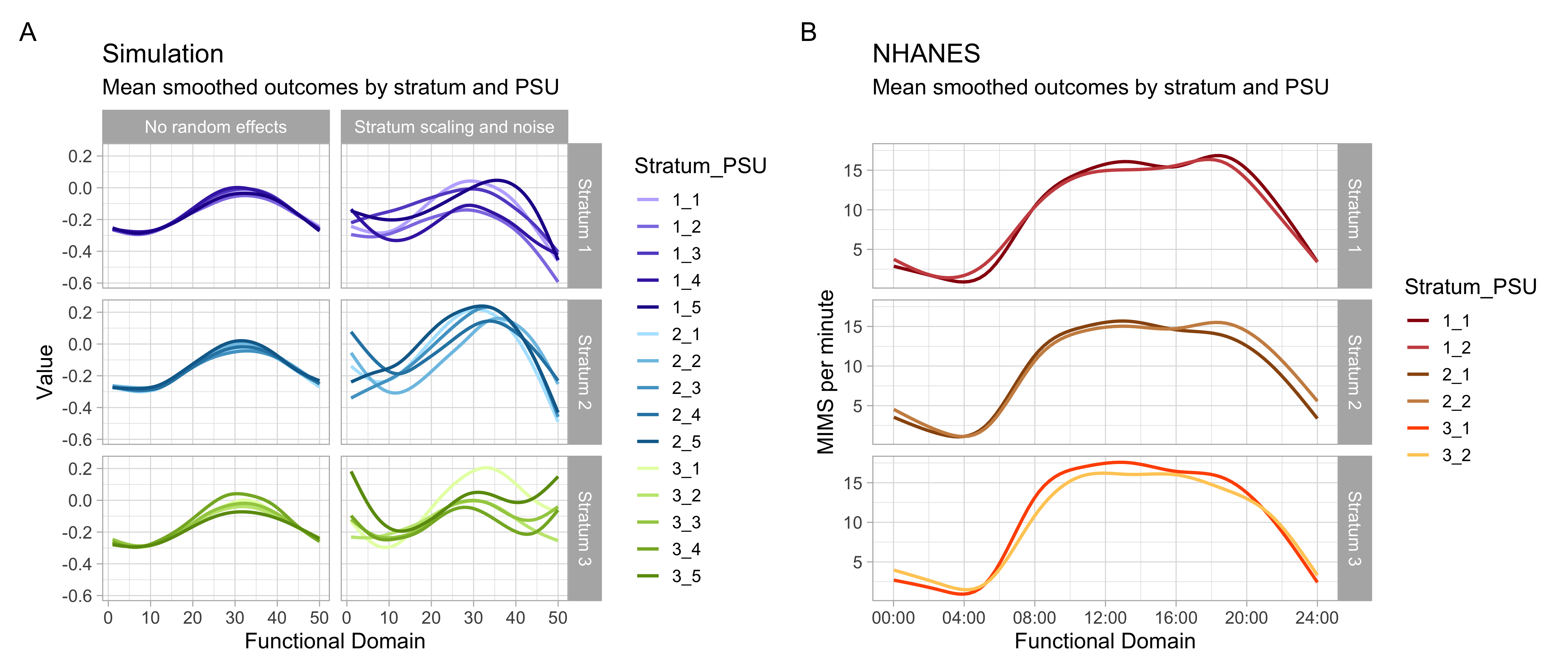}
    \caption{Panel A: smoothed simulated outcomes for three strata (rows) and five PSUs within each stratum (colored lines). Left column: data generated with no random effects. Right column: data generated with strata scaling and random effects. Panel B: smoothed real outcomes for three strata (rows) and the PSUs within each stratum (colored lines) in the NHANES data.}
    \label{fig:sim_outcomes}
\end{figure}

The functional outcomes for each individual in the superpopulation are generated based on this linear predictor. For Gaussian data, we obtain outcomes by sampling from a Normal distribution with mean $\eta_i(s)$ and standard deviation $\sigma^2$, where $\sigma^2$ is varied to account for different values of the signal to noise ratio. For binary data, outcomes are sampled independently from a Bernoulli distribution with probability $p_i(s) = \frac{e^{\eta_i(s)}}{1+e^{\eta_i(s)}}$. For count data, outcomes are sampled independently from a Poisson distribution with mean $e^{\eta_i(s)}$. Panel A of Figure~\ref{fig:sim_outcomes} displays the smoothed outcomes for a few strata and PSUs under different simulation settings. Each row corresponds to a different stratum, and each line correspond to the mean outcome for a PSU within the stratum. The left column corresponds to data generated without random effects, while the right column corresponds to generated data that include strata scaling and PSU-specific random effects. As expected, the right column in Panel A exhibits more heterogeneity than the left column; the right column represents a more realistic data generating mechanism. Indeed, to get a better idea of the level of heterogeneity exhibited by the real NHANES data, Panel B (Figure~\ref{fig:sim_outcomes}) displays displays the average PA measurement for the two PSUs (solid lines) for three strata (each stratum corresponds to one panel in Panel B). 
 
\subsubsection{Theoretical simulation: Sampling}
\begin{figure}
    \centering
    \includegraphics[width=\linewidth]{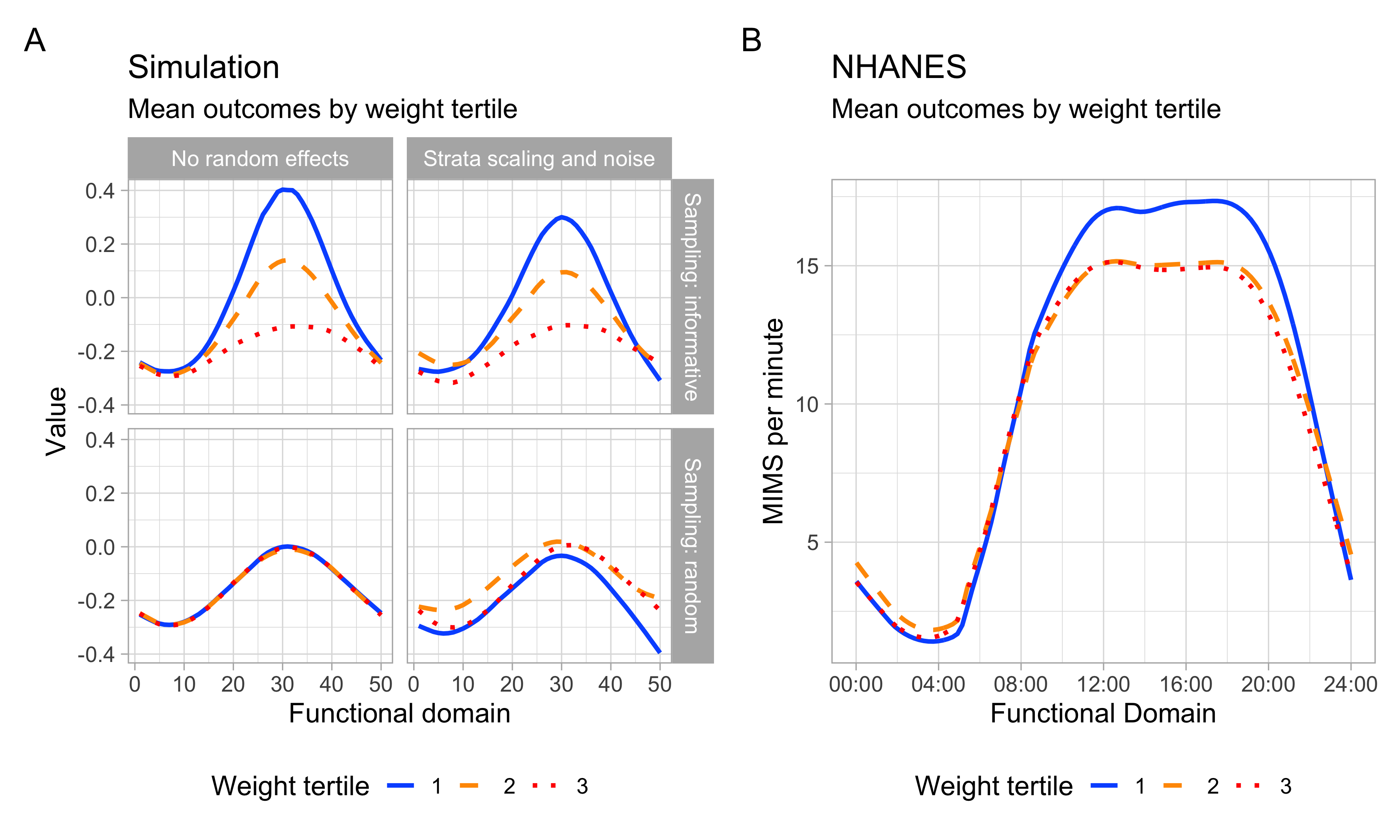}
    \caption{Panel A: mean smoothed outcomes by weight tertile for varied informativeness level and sampling schemes in the simulation. Panel B: mean smoothed outcomes by weight tertile in the NHANES data application. Tertile 1 corresponds to lowest third of weights across the sample, tertile 2 corresponds to middle third of weights in the sample, tertile 3 corresponds to highest third of weights in the sample.}
    \label{fig:weighting}
\end{figure}

Once the superpopulation is generated, many sampling schemes could be used to obtain a sample from it. The survey design dictates how individuals are sampled from the superpopulation. Here, we use two stage stratified sampling. At the first stage, we sample PSUs from each stratum, using probability proportional to size without replacement (PPSWOR) \citep{ppswor}. At the second stage, we sample individuals from each PSU using Poisson sampling. 
More precisely, in stage one, two PSUs are selected from each stratum with the selection probability for each PSU being equal to the ratio of the number of individuals in the PSU to the number of individuals in the stratum. In stage two, individuals are selected from each PSU, in an optionally informative manner. Non-informative sampling is obtained by sampling individuals randomly with the same probability. Informative sampling \citep{parker_bayesian_2023, survey_sofr, informative} is obtained by allowing the selection probability to depend on the mean of their functional outcome: $\pi_i =f( \overline{Y}_i)$. The purpose of simulating informative sampling is to but is to evaluate the robustness of our approach to outcome-related selection bias, which may occur in practice due to differential nonresponse, for example. A higher informativeness level is reflected in (1) higher correlation between survey weight and selection probability and (2) higher coefficient of variation in survey weights. For more information on the informative sampling procedure, see the simulation vignette in Web Appendix C. The second-stage selection probabilities are adjusted to obtain the desired amount of individuals from each PSU, and the selection probabilities are stored at each stage. Panel A of Figure~\ref{fig:weighting} demonstrates the implications of different sampling schemes and random effects structures. In each plot, the mean functional outcome by tertile of survey weight is shown. Large differences in the outcome by survey weight occur for the informative sampling scenarios (top row), there are no discernible differences under random sampling and no random effects (bottom row, left column), and smaller differences occur with random sampling and strata scaling and noise for random effects (bottom row, right column). For comparison, the real NHANES data is shown in Panel B (Figure~\ref{fig:weighting}). While there are no differences in the functional outcome between tertiles 2 and 3, there does appear to be a difference between tertile 1 and tertiles 2 and 3, respectively. 

\subsubsection{Theoretical simulation: Simulation settings}
To evaluate our method, we vary: 
\begin{enumerate}
    \item Simulation parameters
    \begin{itemize}
        \item Distribution of functional responses: (a) Gaussian, (b) Bernoulli, (c) Poisson
    \end{itemize}
    \item Sample size parameters 
    \begin{itemize}
        \item Number of subjects per PSU/strata combination $I_n = \{100, 500\}$
        \item Dimension of the functional domain $L\in\{50, 100, 1440\}$
    \end{itemize}
        \item Relative importance of random effects \citep{FAMM} SNR$_b \in \{ 0.5, 1, 5\}$ where SNR$_b$ is the standard deviation of the fixed effects functions divided by the standard deviation of the random effects functions. A higher value corresponds to stronger fixed effects.
        \item Signal-to-noise ratio: SNR$_\epsilon \in \{ 0.5, 1, 5\}$ (Gaussian responses only): standard deviation of the linear predictors divided by standard deviation of the noise. 
    \item Survey parameters 
    \begin{itemize}
        \item Informativeness level: non-informative (random sampling/uniform weights), medium informativeness, high informativeness (see Web Appendix C for more information)
        \item Strata/PSU random effects: none, correlated stratum/PSU noise only, correlated stratum/PSU noise and stratum slope scaling.
    \end{itemize}

    \item Smoothing parameters: as a sensitivity analysis, we vary the following smoothing parameters for a subset of the above simulation settings. The default smoothing parameters are $\min(L/2, 35)$ knots, low-rank thin plate spline basis, and GCV/Cp for smoothing parameter estimation method.  
    \begin{itemize}
        \item  Number of knots $n_{\text{knots}} \in \{\min(L/4, 20), \min(L/2, 35), \min(3L/4, 50)\}$
        \item Type of smooth: low-rank thin plate spline (\texttt{bs = `tp'}), cubic splines (\texttt{bs = `cr'}), P-splines (\texttt{bs = `ps'})
        \item  Smoothing parameter estimation method: GCV.Cp, REML
    \end{itemize}
\end{enumerate}
We implement the survey-weighted GLM for estimation and  three types of bootstrap for inference: BRR, RWYB, and weighted by the original survey weights. We compare results to the unweighted FUI model with a standard bootstrap of subjects for inference. Methods are compared in terms of accuracy of the functional coefficient estimation and confidence interval coverage. Estimation accuracy was measured by the mean integrated squared error (MISE). The Integrated Squared Error (ISE) is defined as: 
$${\rm ISE}(\widehat{\boldsymbol{\beta}}) = \int_0^1\{\boldsymbol{\widehat{\beta}}(s) - \boldsymbol{\beta}(s)\}^2ds\;,$$
while MISE is the average  ISE over simulations. The true coefficients, $\boldsymbol{\beta}(s)$, are estimated from fitting an unweighted model in the superpopulation.

\subsubsection{Theoretical simulation: Quantifying within-strata and PSU correlation}

To quantify the strength of within-strata/PSU correlation, we use multilevel functional principal components analysis (MFPCA) \citep{mfpca, fda_book}. We define level 1 as strata and PSU membership, and level 2 as individuals within each PSU, and then calculate the proportion of variability explained by the level 1 and level 2 variation, respectively. The higher the proportion variability explained by level 1, the higher the within strata/PSU correlation.

Our simulation settings ranged from $0.3$\% to $86$\% variation explained by strata and PSU. The lowest percent variation explained occurred in the setting with no random effects, uniform sampling, $L = 100$, $I_n = 500$, SNR$_\epsilon$ = 5, and Gaussian data. The highest variation explained occurred with strata/PSU noise, uniform sampling, $L=100$, $I_n=500$, $\text{SNR}_b = 0.5$, and SNR$_\epsilon=5$ with Gaussian data. In general, low variation explained by level 1 occurred in settings with no random effects and uniform sampling. High variation explained by level 1 occurred in settings with high importance of random effects and high signal-to-noise ratios. By comparison, in our data application (see~\ref{subsec:application}), less than 3\% of the variability in the functional outcome is explained by strata and PSU. See Web Appendix D for more on within-strata and PSU correlation; all code for the simulations is publicly available at \url{https://github.com/jhuwit/survey_FoSR}.

\subsubsection{Theoretical simulation: Results}\label{subsec:theosim}
\begin{figure}[ht]
    \centering
     \includegraphics[width=\linewidth]{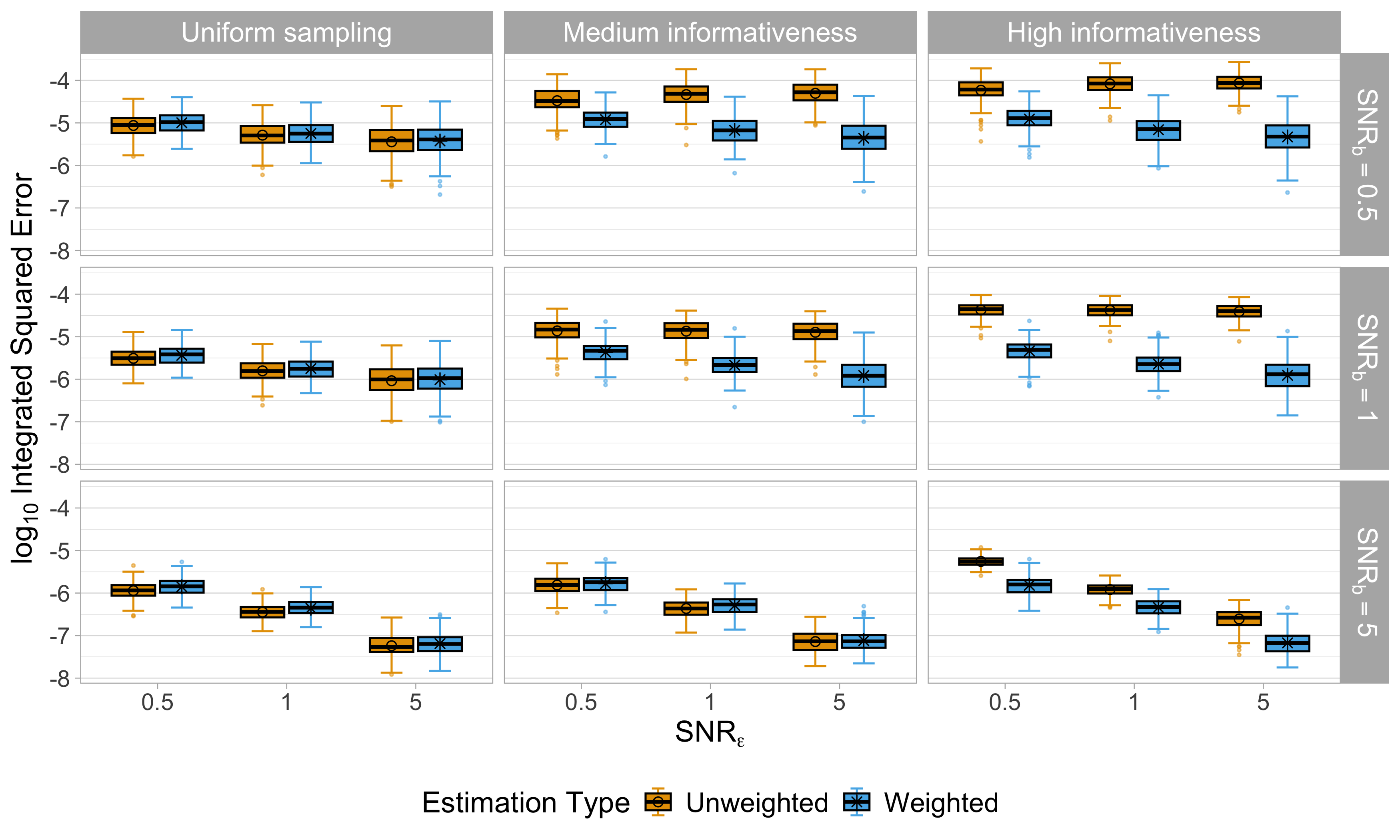}
    \caption{Estimation accuracy for the unweighted (yellow) and weighted (blue) estimation procedure from 200 simulations with varied sampling schemes. Response is Gaussian and the baseline setting is  $I_n=100, L=50$, and stratum scaling and correlated stratum/PSU noise.}
    \label{fig:sampling_res}
\end{figure}

The estimation error for Gaussian data under different sampling schemes for the functional coefficient are shown in Figure~\ref{fig:sampling_res}. Across all panels, $I_n$ and $L$ are fixed at $100$ and $50$, respectively, and there is both strata scaling and strata/PSU random effects. Each facet is a combination of sampling scheme and relative importance of random effects; within each facet, the x-axis represents the signal to noise ratio. Under uniform sampling, MISE is slightly higher for weighted compared to unweighted estimation. For medium informativeness, MISE is lower for weighted compared to unweighted estimation, except when $\text{SNR}_b = 5$, i.e. fixed effects dominate random effects (bottom row, middle panel). At high informativeness, the weighted estimate always achieves lower MISE than the unweighted estimate, even when fixed effects dominate random effects. The difference between the weighted and unweighted estimate is largest for high informativeness, high signal-to-noise ratio, and high relative importance of random effects. Figure S1 in Web Appendix D shows the same  information for the functional intercept. Across all scenarios, for the intercept, the weighted estimate has lower MISE than the unweighted estimate. This is probably due to the presence of the random effects.  

\begin{figure}[ht]
    \centering
    \includegraphics[width=\linewidth]{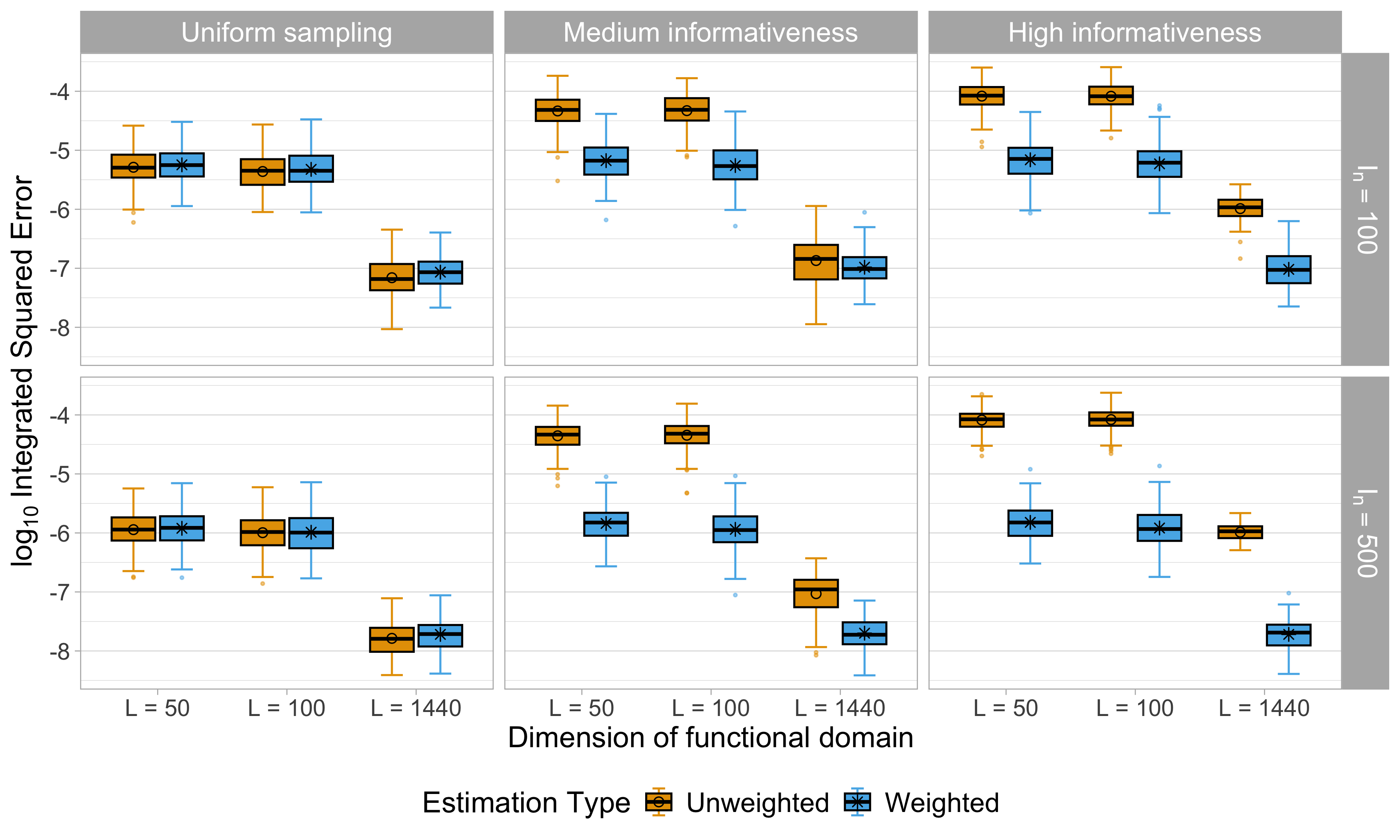}
    \caption{Estimation accuracy for the unweighted (yellow) and weighted (blue) estimation procedure from 200 simulations with varied sample size and dimension of functional domain. Response is Gaussian and the baseline setting is  $\text{SNR}_b=0.5, \text{SNR}_\epsilon = 1$, and stratum scaling and correlated stratum/PSU noise.}
    \label{fig:sample_size}
\end{figure}

Figure~\ref{fig:sample_size} shows MISE for functional coefficient with Gaussian data under different sample size configurations, holding the signal to noise ratio at $1$ and the relative importance of fixed effects at $0.5$. As we saw previously, for uniform sampling, weighted estimation has similar or higher MISE than unweighted estimation; for all other scenarios, weighted estimation has lower MISE. As the dimension of the functional domain increases to $1440$, in the difference between weighted and unweighted estimation shrinks, but remains present, especially for high informativeness sampling. At a given functional domain dimension, differences between weighted and unweighted are more pronounced for $I_n=500$ compared to $I_n=100$. Figure S2 in Web Appendix D shows the same  information for the functional intercept. Again, across all scenarios, the weighted estimate has lower MISE than the unweighted estimate.

\begin{figure}[ht]
    \centering
    \includegraphics[width=\linewidth]{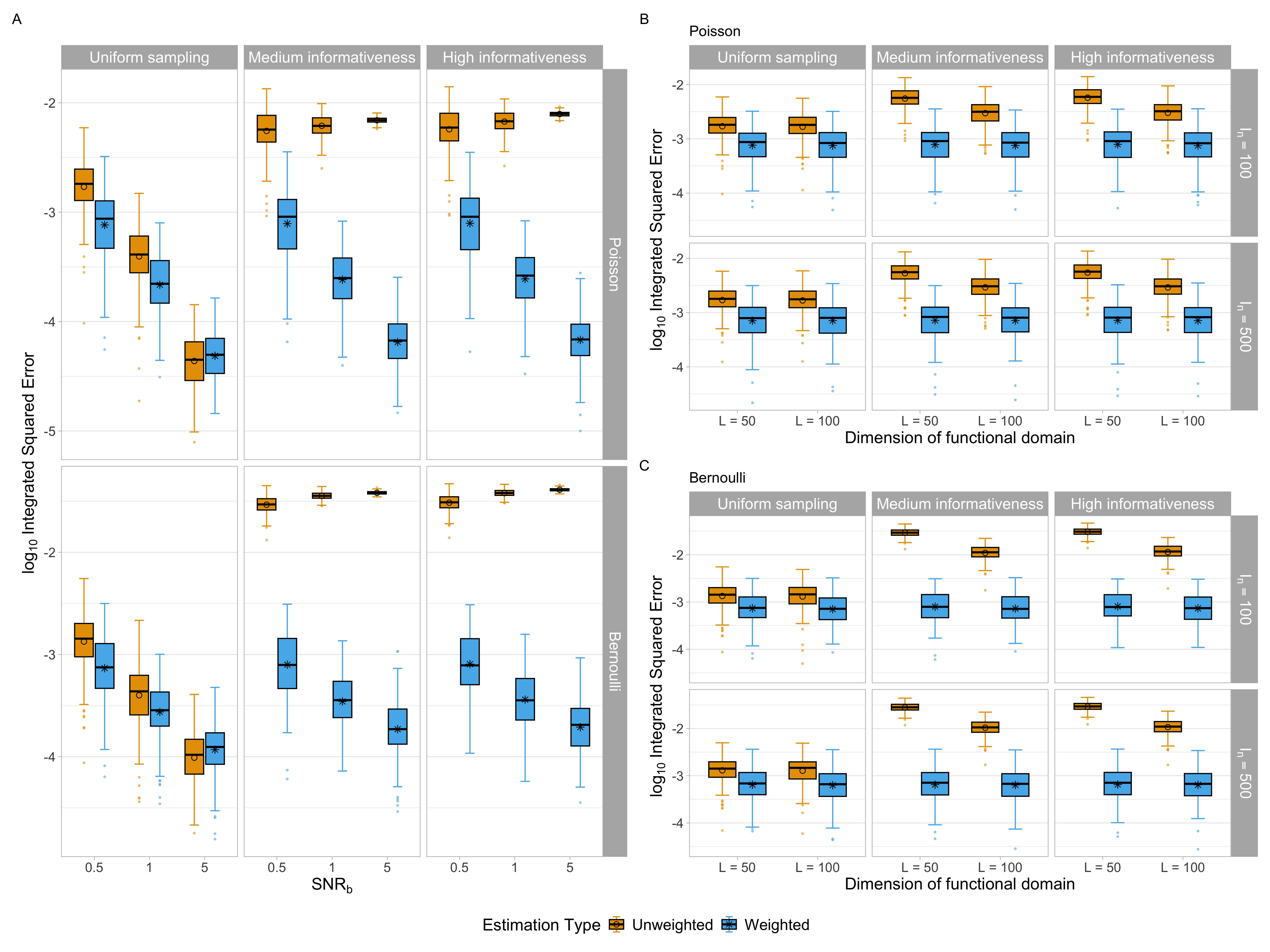}
    \caption{Estimation accuracy for the unweighted (yellow) and weighted (blue) estimation procedure from 200 simulations with varied informativeness and random effect importance (panel A) and varied sample size and dimension of functional domain (panels B and C).}
    \label{fig:res_ng}
\end{figure}

Figure~\ref{fig:res_ng} displays analogous results for the non-Gaussian setting. Panel A demonstrates MISE as a function of sampling scheme, distribution, and relative importance of random effects. Under uniform sampling, weighted estimation has lower MISE than unweighted estimation, except when fixed effects dominate ($\text{SNR}_b =5$). At higher levels of informativeness, weighted estimation has lower MISE than weighted estimation. Panels B and C show the relationship between MISE and sample size parameters for Poisson and Bernoulli data, respectively. Patterns mirror those for Gaussian data.

Table~\ref{tab:pw_cover} displays mean pointwise coverage for the Gaussian data. The baseline setting is Gaussian response, $I_n=100, L=50, \text{SNR}_b=0.5, \text{SNR}_\epsilon = 1$, medium informativeness, and strata scaling with strata/PSU random effects. All other parameters are fixed at their baseline values when another parameter is changed. For the coefficient $X$, under uniform sampling, all methods achieve the nominal ($95$\%) coverage rate. For the coefficient in most non-uniform sampling settings, the unweighted method (range: $0.17$-$0.96$) is much lower than the nominal coverage rate: the exceptions are no random effects (coverage: $0.96$), dominance of fixed effects over random effects (coverage: $0.94$), and high-dimension of the functional domain (coverage: $0.84$). The weighted method (range: $0.94$-$0.97$), BRR (range: $0.94$-$0.96$), and RWYB (range: $0.94$-$0.96$)  achieve nominal or close to nominal coverage in all scenarios. For the intercept, patterns are different. Neither the unweighted (range: $0.09$-$0.40$) nor weighted (range: $0.12$-$0.93$) achieve nominal coverage in \textit{almost any} scenario, even uniform sampling. The exception is when there are no random effects (weighted coverage: $0.93$).  This range of coverage is likely due to the intercept absorbing the random effects shifts from strata and PSUs and the misspecified GLM that does not estimate these effects. Future work will focus on a bias correction procedure. 

Table~\ref{tab:pw_cover_ng} displays mean pointwise coverage for the functional coefficient with non-Gaussian data. The baseline setting is $I_n=100, L=50, \text{SNR}_b=0.5$, medium informativeness, and strata scaling with strata/PSU random effects. All other parameters are fixed at their baseline values when another parameter is changed. The unweighted method fails to achieve the nominal coverage rate in all settings. The weighted method also does not achieve nominal coverage (Bernoulli range: $0.32$-$0.91$, Poisson range: $0.22-0.91$), while RWYB and BRR both achieve nominal or close-to-nominal coverage in all settings (BRR range: $0.93$-$0.95$, RWYB range: $0.93$-$0.95$ for Poisson; BRR range: $0.93$-$0.96$, RWYB range: $0.94$-$0.96$ for Bernoulli). Joint coverage (CMA) results are presented in Web Appendix D. Results have a similar pattern to pointwise coverage: BRR and RWYB outperform unweighted in all settings except for when there are no random effects; however, joint coverage is lower than the nominal rate. For Gaussian data, BRR and RWYB joint coverage ranges from 0.81 to 0.93; for non-Gaussian data, BRR and RWYB joint coverage ranges from 0.81 to 0.95. We hypothesize that lower than expected joint coverage, especially in non-Gaussian data, is due to the smoothing procedure, and future work may focus on solutions to this problem.

Results of the smoothing sensitivity analysis are included in Web Appendix E. Estimation error and coverage were not sensitive to the number of knots, type of smoother, or method for smoothing parameter selection that we tested. Additional simulation results are included in Web Appendix D and the full results of the simulations are available online at \url{https://github.com/jhuwit/survey_FoSR/blob/main/results/simulations/all_survey_res.rds}. 

\begin{footnotesize}
\begin{table}[]
\renewcommand{\arraystretch}{0.75}
    \centering
     \resizebox{\textwidth}{!}{%
    \begin{tabular}{lllllll}
\toprule
\multirow{2}{*}{Model} & \multicolumn{2}{l}{Uniform sampling} &  \multicolumn{2}{l}{Medium informativeness} &  \multicolumn{2}{l}{High informativeness}\\
 & Intercept & $X$ & Intercept & $X$ & Intercept & $X$\\
 \midrule 
Unweighted & \textbf{0.21} & 0.97 &  \textbf{0.21} &  \textbf{0.40} &  \textbf{0.20} &  \textbf{0.22}\\
Weighted & \textbf{0.26} & 0.95 & \textbf{0.26}  & 0.94 & \textbf{0.26} & 0.94\\
BRR & 0.94 & 0.95 & 0.94 & 0.95 & 0.94 & 0.95\\
RWYB & 0.94 & 0.95 & 0.94 & 0.95 & 0.94 & 0.95\\
\midrule
\multirow{2}{*}{Model} &   \multicolumn{2}{l}{No random effects} & \multicolumn{2}{l}{Noise only} & \multicolumn{2}{l}{Scaling \& noise}\\
  & Intercept & $X$ & Intercept & $X$ & Intercept & $X$\\
\midrule 
Unweighted & \textbf{0.32} & 0.96 & \textbf{0.17} & \textbf{0.56} & \textbf{0.21} & \textbf{0.40}\\
Weighted & 0.93 & 0.95 & \textbf{0.26} & 0.95 & \textbf{0.26} & 0.94\\
BRR & 0.95 & 0.94 & 0.95 & 0.96 & 0.94 & 0.95\\
RWYB & 0.95 & 0.95 & 0.94 & 0.96 & 0.94 & 0.95\\
\midrule 
\multirow{2}{*}{Model}  & \multicolumn{2}{l}{$\text{SNR}_b = 0.5$} & \multicolumn{2}{l}{$\text{SNR}_b=1$} &  \multicolumn{2}{l}{$\text{SNR}_b = 5$}\\
 & Intercept & $X$ & Intercept & $X$ & Intercept & $X$\\
 \midrule 
Unweighted & \textbf{0.21} & \textbf{0.40} & \textbf{0.23} & \textbf{0.34} & \textbf{0.40} & 0.94\\
Weighted & \textbf{0.26} & 0.94 & \textbf{0.28} & 0.95 & \textbf{0.59} & 0.94\\
BRR & 0.94 & 0.95 & 0.95 & 0.95 & 0.95 & 0.94\\
RWYB & 0.94 & 0.95 & 0.94 & 0.95 & 0.95 & 0.95\\
\midrule 
\multirow{2}{*}{Model}  & \multicolumn{2}{l}{$\text{SNR}_\epsilon = 0.5$} & \multicolumn{2}{l}{$\text{SNR}_\epsilon=1$} &  \multicolumn{2}{l}{$\text{SNR}_\epsilon = 5$}\\
 & Intercept & $X$ & Intercept & $X$ & Intercept & $X$\\
 \midrule 
Unweighted & \textbf{0.21} & \textbf{0.65} & \textbf{0.21} & \textbf{0.40} & \textbf{0.18} & \textbf{0.33}\\
Weighted & \textbf{0.33} & 0.94 & \textbf{0.26} & 0.94 & \textbf{0.22} & 0.94\\
BRR & 0.95 & 0.95 & 0.94 & 0.95 & 0.94 & 0.95\\
RWYB & 0.95 & 0.96 & 0.94 & 0.95 & 0.94 & 0.95\\
\midrule
\multirow{2}{*}{Model} &  \multicolumn{2}{l}{$I_n=100$}& \multicolumn{2}{l}{$I_n=500$} & \multicolumn{2}{l}{}\\
& Intercept & $X$ & Intercept & $X$ &  & \\
\midrule 
Unweighted & \textbf{0.21} & \textbf{0.40} & \textbf{0.09} & \textbf{0.17} &  & \\
Weighted & \textbf{0.26} & 0.94 & \textbf{0.12} & 0.95 &  & \\
BRR & 0.94 & 0.95 & 0.94 & 0.95 &  & \\
RWYB & 0.94 & 0.95 & 0.94 & 0.95 &  & \\
\midrule 
\multirow{2}{*}{Model} & \multicolumn{2}{l}{$L=50$}& \multicolumn{2}{l}{$L=100$} & \multicolumn{2}{l}{$L=1440$}\\
 & Intercept & $X$ & Intercept & $X$ & Intercept & $X$\\
\midrule 
Unweighted & \textbf{0.21} & \textbf{0.40} & \textbf{0.20} & \textbf{0.38} &  & \textbf{0.84}\\
Weighted & \textbf{0.26} & 0.94 & \textbf{0.24} & 0.94 & & 0.97\\
BRR & 0.94 & 0.95 & 0.94 & 0.95 &  & 0.95\\
RWYB & 0.94 & 0.95 & 0.94 & 0.95 &  & 0.96\\
\bottomrule
\end{tabular}}
    \caption{Gaussian data: empirical coverage probability of 95\% pointwise coverage bands with each method from 200 simulations. The pointwise band is calculated as mean$\pm$2 SD. The baseline setting is $I_n=100, L=50, \text{SNR}_b=0.5, \text{SNR}_\epsilon = 1$, medium informativeness, and stratum scaling and correlated stratum/PSU noise. All other parameters are fixed at their baseline values when another parameter is changed. Values less than 0.93 are bolded to highlight settings where a method fails to achieve nominal coverage. The intercept coverage for $L=1440$ is not presented, as computing the true intercept for this dimension of functional domain was infeasible due to memory limitations.}
    \label{tab:pw_cover}
\end{table}
\end{footnotesize}

\begin{table}[]
\renewcommand{\arraystretch}{0.75}
    \centering
     \resizebox{\textwidth}{!}{%
    \begin{tabular}{lllllll}
\toprule
\multirow{2}{*}{Model} & \multicolumn{2}{l}{Uniform sampling} &  \multicolumn{2}{l}{Medium informativeness} &  \multicolumn{2}{l}{High informativeness}\\
 & Poisson & Bernoulli &  Poisson & Bernoulli & Poisson & Bernoulli\\
 \midrule 
Unweighted & \textbf{0.31} & \textbf{0.41} & \textbf{0.11} & \textbf{0.00} & \textbf{0.11} & \textbf{0.00}\\
Weighted & \textbf{0.38} & \textbf{0.55} & \textbf{0.41} & \textbf{0.58} & \textbf{0.40} & \textbf{0.58}\\
BRR & 0.93 & 0.94 & 0.93 & 0.93 & 0.93 & 0.94\\
RWYB & 0.93 & 0.94 & 0.93 & 0.94 & 0.93 & 0.94\\
\midrule
\multirow{2}{*}{Model} &   \multicolumn{2}{l}{No random effects} & \multicolumn{2}{l}{Noise only} & \multicolumn{2}{l}{Scaling \& noise}\\
  & Poisson & Bernoulli &  Poisson & Bernoulli & Poisson & Bernoulli\\
\midrule 
Unweighted & \textbf{0.00} & \textbf{0.00} & \textbf{0.16} & \textbf{0.00} & \textbf{0.11} & \textbf{0.00}\\
Weighted & \textbf{0.91} & \textbf{0.90} & \textbf{0.41} & \textbf{0.60} & \textbf{0.41} & \textbf{0.58}\\
BRR & 0.95 & 0.95 & 0.93 & 0.94 & 0.93 & 0.93\\
RWYB & 0.95 & 0.95 & 0.93 & 0.94 & 0.93 & 0.94\\
\midrule 
\multirow{2}{*}{Model}  & \multicolumn{2}{l}{$\text{SNR}_b = 0.5$} & \multicolumn{2}{l}{$\text{SNR}_b=1$} &  \multicolumn{2}{l}{$\text{SNR}_b = 5$}\\
 & Poisson & Bernoulli &  Poisson & Bernoulli & Poisson & Bernoulli\\
 \midrule 
Unweighted & \textbf{0.11} & \textbf{0.00} & \textbf{0.00} & \textbf{0.00} & \textbf{0.00} & \textbf{0.00}\\
Weighted & \textbf{0.41} & \textbf{0.58} & \textbf{0.60} & \textbf{0.76} & \textbf{0.89} & \textbf{0.91}\\
BRR & 0.93 & 0.93 & 0.93 & 0.94 & 0.94 & 0.96\\
RWYB & 0.93 & 0.94 & 0.93 & 0.94 & 0.95 & 0.96\\
\midrule 
\multirow{2}{*}{Model} &  \multicolumn{2}{l}{$I_n=100$}& \multicolumn{2}{l}{$I_n=500$} & \multicolumn{2}{l}{}\\
 & Poisson & Bernoulli &  Poisson & Bernoulli &  &\\
\midrule 
Unweighted & \textbf{0.11} & \textbf{0.00} & \textbf{0.06} & \textbf{0.00} &  & \\
Weighted & \textbf{0.41} & \textbf{0.58} & \textbf{0.22} & \textbf{0.32} &  & \\
BRR & 0.93 & 0.93 & 0.93 & 0.94 &  & \\
RWYB & 0.93 & 0.94 & 0.93 & 0.94 &  & \\
\midrule 
\multirow{2}{*}{Model} & \multicolumn{2}{l}{$L=50$}& \multicolumn{2}{l}{$L=100$} & \multicolumn{2}{l}{}\\
 & Poisson & Bernoulli &  Poisson & Bernoulli & & \\
\midrule 
Unweighted & \textbf{0.11} & \textbf{0.00} & \textbf{0.19} & \textbf{0.02} &  & \\
Weighted & \textbf{0.41} & \textbf{0.58} & \textbf{0.33} & \textbf{0.47} & & \\
BRR & 0.93 & 0.93 & 0.93 & 0.94 & & \\
RWYB & 0.93 & 0.94 & 0.94 & 0.94 & & \\

\bottomrule
\end{tabular}}
    \caption{Non-Gaussian data: empirical coverage probability of 95\% pointwise coverage bands for coefficient $X$ with each method from 200 simulations. The pointwise band is calculated as mean$\pm$2 SD. $I_n=100, L=50, \text{SNR}_b=0.5$, medium informativeness, and stratum scaling and correlated stratum/PSU noise. All other parameters are fixed at their baseline values when another parameter is changed. Values less than 0.93 are bolded to highlight settings where a method fails to achieve nominal coverage.}
    \label{tab:pw_cover_ng}
\end{table}

\subsection{Empirical simulation}\label{subsec:empirical}
The goal of the empirical simulation is to confirm that the approach can adjust for informative sampling in the context of estimating the relationship between scalar covariates and a physical activity functional response curve. The structure for the empirical simulation is modeled after Parker and Holan's NHANES simulation \citep{parker_bayesian_2023}.

To create this simulation, we treat the entire NHANES sample data as our population, and subsample from this population informatively. To do so, we first calculate two variables: (1) the individual's original survey weight $w_i$, scaled to have mean $0$ and variance $1$ across the population, which we denote $w_i^{sc}$, and (2) the individual's mean physical activity measured in Monitor Independent Movement Summary (MIMS) units \citep{mims}, also scaled to have mean $0$ and variance $1$, which we denote $m_i^{sc}$. We truncate the weights at $-2$ and $2$ to prevent extreme weight values. Then we consider four sampling scenarios/weighting schemes. Let $\text{logistic}(x) = \frac{1}{1+e^{-x}}$ denote the logistic transformation.
\begin{itemize}
    \item Uniform sampling: $\pi_i \propto \frac{1}{n}$ (uniform)
    \item Sampling probability proportional to the individual's scaled survey weight: $\pi_i \propto \text{logistic}(w_i^{sc})$ (NHANES)
    \item Sampling probability proportional to the individual's scaled physical activity: $\pi_i \propto \text{logistic}(m_i^{sc})$ (MIMS)
    \item Sampling probability proportional to combination of individual's scaled physical activity and survey weight:  $\pi_i \propto \text{logistic}(0.5w_i^{sc} + 0.5m_i^{sc})$ (NHANES\&MIMS)
\end{itemize}

We scale the $\pi_i$ to ensure the desired amount of individuals selected in the subsample (we chose $n=1000$) and set $w_i = 1/\pi_i$ as the survey weight for each individual in the subsample.

The functional response is the mean MIMS profile over the course of the day and the scalar covariate is sex. We fit three models: an unweighted FoSR model with unweighted bootstrap for inference, the survey weighted FoSR model with survey weighted bootstrap for inference, and the survey weighted FoSR model with BRR for inference. We do not use RWYB because we do not have the required selection probabilities at each sampling stage. From each model we obtain a functional coefficient estimate, and for each iteration, we calculate the MISE by comparing the functional coefficient in the subsample to the (unweighted) functional coefficient in the entire population. We also calculate the pointwise and joint coverage, again comparing to the functional coefficient in the entire population. We repeat the simulation $200$ times. The ground truth in this simulation is not a data generating mechanism, but the estimated effect from an FUI model fit on the entire NHANES dataset.

\subsubsection{Empirical simulation: Results}
\begin{figure}[!ht]
    \centering
    \includegraphics[width=\textwidth]{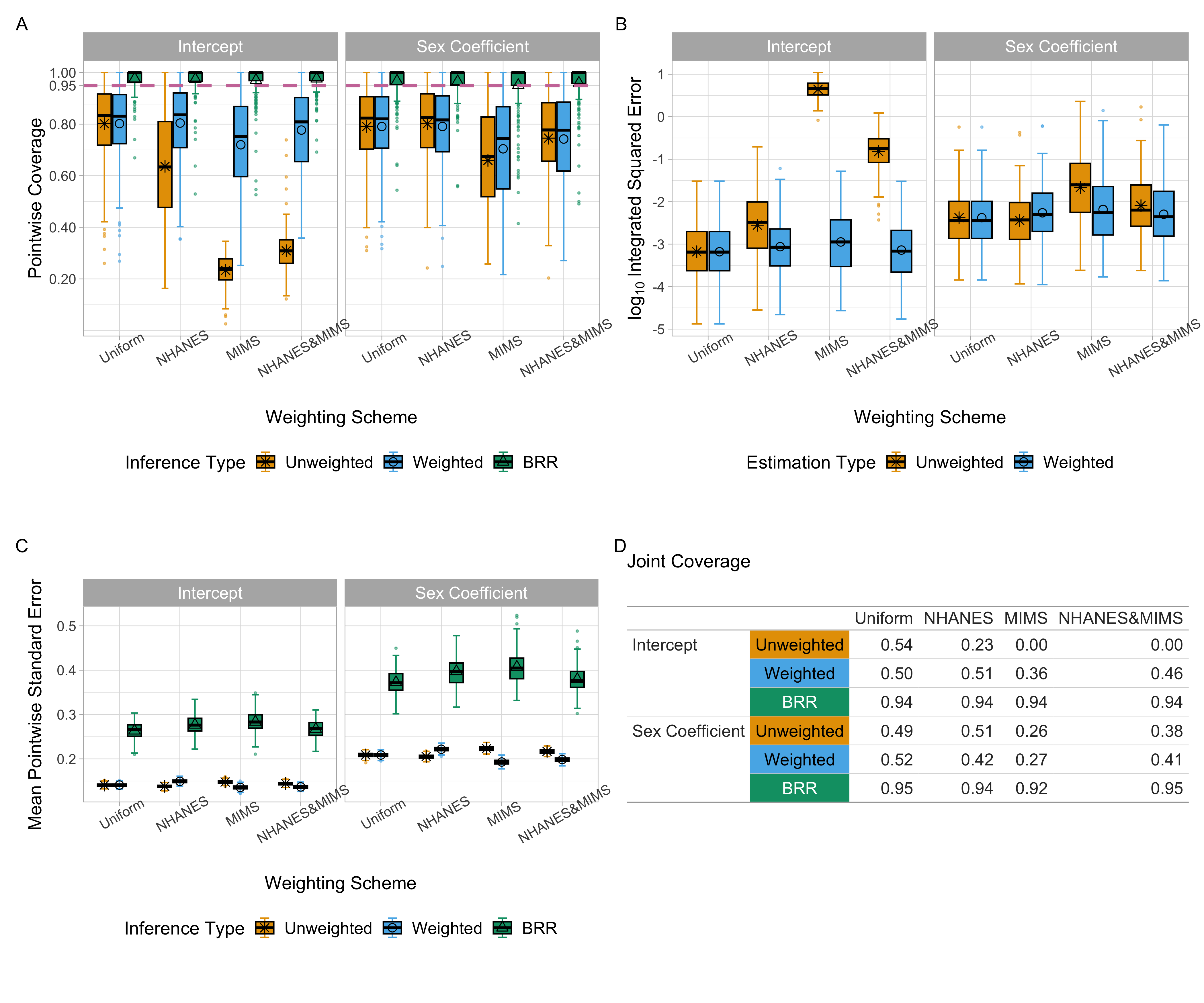}
    \caption{Empirical simulation results. The x-axis shows the four weighting schemes used: uniform weighting, weights based on NHANES weights, weights based on physical activity profiles (MIMS), and weights based on a combination of NHANES weights and MIMS). Panel A: pointwise coverage; Panel B: log MISE; Panel C: mean pointwise standard error. Panel D contains a table with mean joint coverage across all iterations.}
    \label{fig:emp_sim}
\end{figure}

Figure~\ref{fig:emp_sim} displays the results of the empirical simulation. Panels A, B, C, and D show the pointwise coverage, log MISE, mean pointwise standard error, and joint coverage across all 200 iterations for the three models: unweighted, weighted, and BRR, while the weighting schemes (uniform, NHANES, MIMS, NHANES\&MIMS) are displayed on the x-axis. In the uniform weighting scheme, all methods achieve adequate coverage, and pointwise standard errors are similar across methods. In the NHANES weighting scheme, the unweighted method has lower coverage and higher MISE for $\beta_0(s)$, but not for $\beta_1(s)$. In the MIMS weighting scheme, the unweighted method has low coverage and high MISE for both $\beta_0(s)$, but not for $\beta_1(s)$, and the weighted method has slightly lower than nominal coverage (pointwise coverage = $0.94$, joint coverage = $0.92$) for $\beta_1(s)$. The  NHANES\&MIMS scenario is a weighted average of the NHANES and MIMS results. Across all weighting schemes, the weighted and BRR methods achieve similar coverage rates, although the pointwise standard error for BRR is typically larger than that of weighted. In this setting, since we do not create or observe high within PSU/strata correlation, the survey weighted method performs relatively well. 

\section{Application}\label{subsec:application}
We apply the approach to the NHANES data discussed in the introduction. NHANES is a large epidemiological study carried out in two-year waves to measure the health and nutritional status of Americans. In 2011-2014, participants were given a wrist-worn accelerometer to wear continuously for up to seven full days. A measure of physical activity, Monitor Independent Movement Summary (MIMS) units \citep{mims}, are calculated from the raw data and provided by NHANES at one minute resolution \citep{paxming, paxminh}. Here we focus on individuals 18 and older at the time of wearing the device. We define a valid day as at least 1368 minutes classified as wear and without data quality issues (95\% of a full day); we exclude individuals with less than three valid days. To obtain one function per participant, we average the MIMS observations over minutes from valid days. The final sample includes 8664 individuals with 1440 observations per day. Here, we are interested in investigating the relationship between MIMS as outcomes measured at every minute of the day and scalar predictors (covariates).  We use the following notation: $X_{i1}$ is a binary indicator of sex (female = 1), $X_{i2}$ is a binary indicator of age $\in [30,49]$, $X_{i3}$ is a binary indicator of age $\in [50,64]$, $X_{i4}$ is a binary indicator of age $\geq 65$ (such that the reference category for age is $18-29$). Age is categorized to capture nonlinear effects and preserve interpretability. Finally, $X_{i5}$ is a binary indicator of Hispanic ethnicity, $X_{i6}$ is a binary indicator of non-Hispanic Black race/ethnicity, and $X_{i7}$ is a binary indicator of ``other'' race/ethnicity category, such that the reference category for race/ethnicity is non-Hispanic white. 

A survey-weighted GLM is fit separately at all 1440 locations (minutes of the day). The pointwise coefficient estimates are smoothed, and two bootstrap methods (weighted and BRR) are applied. RWYB is not used because the requisite sampling probabilities at each stage are not provided by NHANES. In addition, the unweighted model with an unweighted bootstrap is fit for comparison. To account for the fact each participant's mean physical activity may be estimated from between three and seven days of wear, we investigate incorporating precision weighting in the pointwise models, giving larger weights to participants who have seven full days.

\begin{figure}[!ht]
    \centering
    \includegraphics[width=\linewidth]{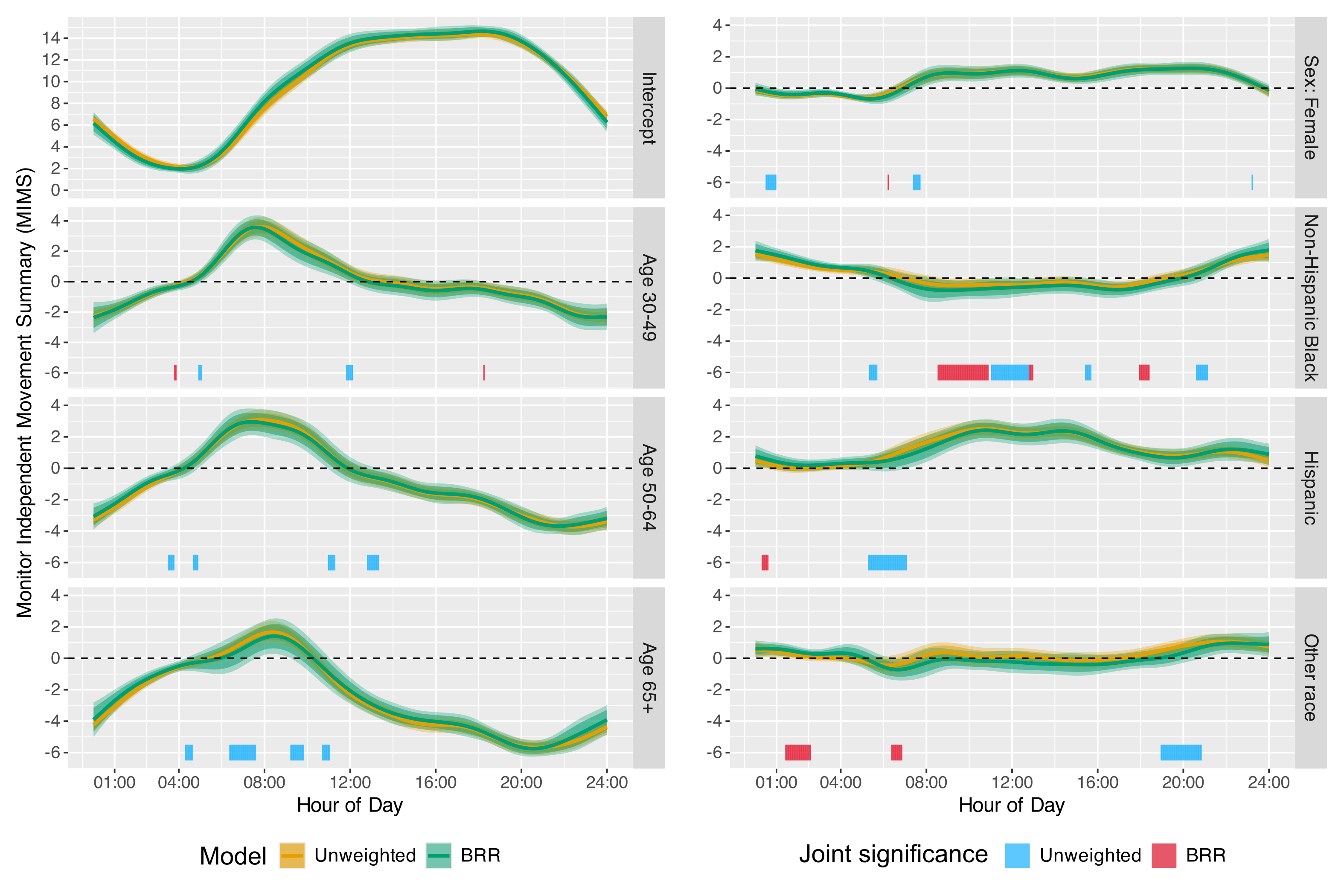}
    \caption{Estimated functional intercept and coefficients along with their associated pointwise (shaded) and joint (lighter shading) confidence bands from unweighted (orange) and BRR (green) regression models. The reference categories are male, age 18-29, and white non-Hispanic race. The bars at the bottom of each facet indicate areas where the joint confidence bands for unweighted and BRR disagree: blue bars indicate unweighted is significant and BRR is not while red bars indicate BRR is significant while unweighted is not.}
    \label{fig:application_width}
\end{figure}

Figure~\ref{fig:application_width} displays the estimated functional coefficients for the intercept along with age, sex, and race categories, along with the associated pointwise and joint confidence bands. The intercept represents the mean MIMS accumulated at each minute across the day for a white non-Hispanic male aged 18-29: for example, the mean MIMS accumulated at 12:00 is approximately 14. The other coefficients represent the mean number of additional MIMS for each demographic group compared to the reference group, holding the other covariates constant. The coefficient for sex shows that females have higher physical activity than males at all times except for between midnight and 6AM. The age patterns show that all age groups are more active than 18-29 year-olds between 6AM and 10AM, and less active between 10PM and 2AM. Individuals 65 and older are less active after 11AM than 18-29 year-olds. The race/ethnicity patterns show that non-Hispanic Black individuals are more active than non-Hispanic white individuals between 10PM and 4AM, while Hispanic individuals are more active during the daytime hours than non-Hispanic white individuals. The bars at the bottom of each facet of Figure~\ref{fig:application_width} denote areas where the joint confidence bands for unweighted and BRR disagree in significance: blue bars indicate unweighted is significant while BRR is not and red bars indicate where BRR is significant and unweighted is not. While the methods agree most of the time, there are some differences, especially for the race covariates. The BRR confidence intervals are wider than unweighted confidence intervals nearly always: the percent difference ranges from -9\% to 130\%, and the mean is 38\%. The weighted confidence intervals are usually larger than the unweighted confidence intervals: percent difference ranges from -35\% to 112\%, and the mean is 20\%. Interestingly, for the Other race category, the weighted confidence intervals are always narrower than the unweighted confidence intervals, perhaps due to the fact that ``Other'' race has the lowest standard deviation in survey weights across the age, sex, and race categories. The difference between the point estimates for weighted vs. unweighted estimation is small (see Figure S13 in Web Appendix D) and the difference in inferences between the three methods is also small (i.e., areas of significance are similar across methods). For the age and sex categories, all three methods (BRR, weighted, unweighted) agree on significance between 91\% and 100\% of the time; for race categories, they agree between 77\% and 89\% of the time (see Supplemental Table S5 for more details). The results in our application are likely similar across methods due to the low within strata/PSU correlation for physical activity in NHANES. This indicates that previously published MIMS results for these NHANES waves using unweighted methods likely remain valid for covariates with low within-PSU correlation. The incorporation of precision weights has negligible impact on our results (see Figure S15 in Web Appendix D).

\section{Discussion}\label{sec:disc} 
We propose estimation and inference procedures for function on scalar regression under complex survey designs. The approach builds on the FUI approach for function on scalar regression by incorporating survey information into the massively univariate models for estimation, and modifying the bootstrap procedure to be survey-aware for inference. Our approach is implemented in the \texttt{svyfosr} R package. Our analytical and empirical simulation studies demonstrate that our approach accurately captures the shape of the functional coefficients and achieves the nominal 95\% coverage rate for pointwise confidence intervals; joint coverage is slightly lower than the nominal rate. In estimation, our approach is more accurate than the unweighted model under informative sampling, and for inference, our approach outperforms the unweighted bootstrap.

To our knowledge, this is the first approach that incorporates complex survey information into function on scalar regression. Our approach allows for valid estimation and inference in function on scalar regression with exponential family outcomes for a broad class of potential survey designs: BRR and RWYB options allow the user to specify any type of survey design (stratified, two stage, etc.), potentially with different sampling schemes at each stage. While previous work at the intersection of functional data analysis and survey weighting has focused on functional principal components analysis \citep{survey_fpca}, smoothing parameter selection for penalized spline estimation \citep{splines1, splines2, splines3}, and scalar on function regression \citep{survey_sofr, parker_bayesian_2023}, this is the first approach to allow for survey weighting in function on scalar regression. Additionally, we provide R software that is ready to use and uses similar syntax to that of existing FoSR and survey software. 

We also designed a first-of-its kind analytical simulation that (1) generates a superpopulation of functional outcomes that are correlated within strata and PSU and (2) samples informatively from this superpopulation. Our simulations and application to NHANES demonstrate that incorporating survey information is important to provide unbiased estimators and nominal coverage probabilities for the confidence intervals. Our results also clarified that the weighted bootstrap can perform as well as BRR and RWYB methods when there is  low within-strata/PSU correlation, though it can substantially underperform in cases with higher within-strata/PSU correlation (see Figure S12 in Web Appendix D). In NHANES, large differences were not observed between the survey-weighted and unweighted functional coefficients, although confidence bands were wider under the BRR inferential procedure. This difference may be due to the fact that in NHANES, less than 3\% of the variability in MIMS is explained by PSU/strata, i.e. there is low correlation between individuals within the same PSU and stratum. However, when there correlation between PSU and stratum is higher, differences between survey-aware and survey-naive approaches become more pronounced, and it is critical to incorporate survey structure into estimation and inference procedures. 

Our approach has a few limitations and potential extensions. First, while FUI was originally developed for longitudinal (multilevel) data, our approach currently is only implemented for single-level functional data (one function per participant). Survey-aware estimation and inference in linear mixed effects models is not straightforward, although there have been recent methodological developments in the area \citep{survey_mixed}. A future direction is to extend our FUI-based approach to incorporate multilevel survey data. Challenges include computation (Lumley and Huang's method uses profile weighted pairwise deviance, which is not feasible to repeat at each point along the functional domain), and specifying by-stage sampling probabilities: multiday data is different from typical multilevel survey data where sampling levels match the nested structure of the data and it is unclear how the days should be weighted. Second, the strata and PSU level random effects structure in the analytical simulation means that there is no fixed, theoretical population-level intercept function as the target for inference; $\beta_0(s)$ varies across strata and PSU. In our simulation, we estimated the intercept function empirically; however, future work could entail finding a closed-form solution for $\beta_0(s)$ in the superpopulation. Performance of our approach was lowest for non-Gaussian outcomes: minimum pointwise coverage for Bernoulli outcomes with BRR and RWYB inference was 0.90, and MISE for weighted estimation was not always lower than MISE for unweighted estimation in some informative sampling settings. This is likely due to the mean-variance dependence and nonlinearity of the link function for non-Gaussian data in combination with using pointwise estimation. Future extensions on bias correction for non-Gaussian data or joint modeling of the functional outcome may improve performance. Finally, our conclusions about differences in MIMS profiles from the data application are representative of the US population conditional on meeting the wear criterion (three valid days of wear). If individuals who did not meet this wear criterion are systematically different than those who did, our results do not generalize to these individuals.

\section*{Acknowledgments, Funding, Data Availability, Conflicts of Interest}

{\bf Acknowledgments}: The authors would like to thank Dr. Erjia Cui for developing the FUI method and the authors of the R packages \texttt{survey} and \texttt{svrep} for creating and maintaining software for survey-aware analyses.

\noindent {\bf Funding}: This work was supported by the National Institutes of Health under Grants R01NS060910 and R01AG075883. 

\noindent {\bf Data Availability}: All data and code is publicly available at \url{https://github.com/jhuwit/survey_FoSR}

\noindent {\bf Conflicts of Interest}: The authors have no conflicts of interest to report.

\bibliographystyle{oup-plain}
\bibliography{references.bib}

@Manual{R,
    title = {R: A Language and Environment for Statistical Computing},
    author = {{R Core Team}},
    organization = {R Foundation for Statistical Computing},
    address = {Vienna, Austria},
    year = {2023},
    url = {https://www.R-project.org/},
  }

@article{surveys_intro,
  title = {{Introduction to the Design and Analysis of Complex Survey Data}},
  volume = {32},
  ISSN = {0883-4237},
  url = {http://dx.doi.org/10.1214/17-STS614},
  DOI = {10.1214/17-sts614},
  number = {2},
  journal = {Statistical Science},
  publisher = {Institute of Mathematical Statistics},
  author = {Skinner,  Chris and Wakefield,  Jon},
  year = {2017},
  month = may 
}

@techreport{nhanes_guidelines_2018,
  title        = {National Health and Nutrition Examination Survey: Analytic Guidelines, 2011–2016},
  author       = {{National Center for Health Statistics}},
  year         = {2018},
  institution  = {Centers for Disease Control and Prevention},
  address      = {Hyattsville, MD},
  note         = {Available from \url{https://wwwn.cdc.gov/nchs/data/nhanes/analyticguidelines/11-16-analytic-guidelines.pdf}},
  month        = dec,
  type         = {Technical Report}
}

@article{raowuyue1992,
	title = {Some recent work on resampling methods for complex surveys},
	volume = {18},
	pages = {209-217},
	number = {2},
	journal = {Survey Methodology},
	author = {Rao, J.N.K. and Wu, C.F.J. and  Yue, K.},
	year={1992}
}

@article{FAMM,
  title = {{Functional Additive Mixed Models}},
  volume = {24},
  ISSN = {1537-2715},
  url = {http://dx.doi.org/10.1080/10618600.2014.901914},
  DOI = {10.1080/10618600.2014.901914},
  number = {2},
  journal = {Journal of Computational and Graphical Statistics},
  publisher = {Informa UK Limited},
  author = {Scheipl,  Fabian and Staicu,  Ana-Maria and Greven,  Sonja},
  year = {2015},
  month = apr,
  pages = {477–501}
}

@article{ivanescu_2015,
	author = {Ivanescu, Andrada E. and Staicu, Ana-Maria and Scheipl, Fabian and Greven, Sonja},
	journal = {Computational Statistics},
	number = {2},
	pages = {539--568},
	title = {Penalized function-on-function regression},
	volume = {30},
	year = {2015}
}

@article{survey_fpca,
  title = {Properties of design-based functional principal components analysis},
  volume = {140},
  ISSN = {0378-3758},
  url = {http://dx.doi.org/10.1016/j.jspi.2009.06.012},
  DOI = {10.1016/j.jspi.2009.06.012},
  number = {1},
  journal = {Journal of Statistical Planning and Inference},
  publisher = {Elsevier BV},
  author = {Cardot,  Hervé and Chaouch,  Mohamed and Goga,  Camelia and Labruère,  Catherine},
  year = {2010},
  month = jan,
  pages = {75–91}
}

@incollection{gentle1998qr,
  author    = {Gentle, James E.},
  title     = {QR Factorization},
  booktitle = {Numerical Linear Algebra for Applications in Statistics},
  publisher = {Springer-Verlag},
  address   = {Berlin},
  year      = {1998},
  pages     = {95--97},
  chapter   = {3.2.2}
}

@incollection{informative,
title = {Chapter 39 - Inference under Informative Sampling},
editor = {C.R. Rao},
series = {Handbook of Statistics},
publisher = {Elsevier},
volume = {29},
pages = {455-487},
year = {2009},
booktitle = {Handbook of Statistics},
issn = {0169-7161},
doi = {https://doi.org/10.1016/S0169-7161(09)00239-9},
url = {https://www.sciencedirect.com/science/article/pii/S0169716109002399},
author = {Danny Pfeffermann and Michail Sverchkov},
}

@article{splines1,
  title = {Model-assisted estimation for complex surveys using penalised splines},
  volume = {92},
  ISSN = {0006-3444},
  url = {http://dx.doi.org/10.1093/biomet/92.4.831},
  DOI = {10.1093/biomet/92.4.831},
  number = {4},
  journal = {Biometrika},
  publisher = {Oxford University Press (OUP)},
  author = {Breidt,  F. J. and Claeskens,  G. and Opsomer,  J. D.},
  year = {2005},
  month = dec,
  pages = {831–846}
}

@article{splines2,
  title = {Nonparametric regression estimators in complex surveys},
  volume = {85},
  ISSN = {1563-5163},
  url = {http://dx.doi.org/10.1080/00949655.2013.860139},
  DOI = {10.1080/00949655.2013.860139},
  number = {5},
  journal = {Journal of Statistical Computation and Simulation},
  publisher = {Informa UK Limited},
  author = {Zhang,  Guoyi and Christensen,  Fletcher and Zheng,  Wei},
  year = {2013},
  month = nov,
  pages = {1026–1034}
}

@article{splines3,
  title = {Survey design asymptotics for the model-assisted penalised spline regression estimator},
  volume = {25},
  ISSN = {1029-0311},
  url = {http://dx.doi.org/10.1080/10485252.2013.780057},
  DOI = {10.1080/10485252.2013.780057},
  number = {3},
  journal = {Journal of Nonparametric Statistics},
  publisher = {Informa UK Limited},
  author = {McConville,  K. S. and Breidt,  F. J.},
  year = {2013},
  month = sep,
  pages = {745–763}
}

@article{Xiao2019,
  title = {Asymptotic theory of penalized splines},
  volume = {13},
  ISSN = {1935-7524},
  url = {http://dx.doi.org/10.1214/19-EJS1541},
  DOI = {10.1214/19-ejs1541},
  number = {1},
  journal = {Electronic Journal of Statistics},
  publisher = {Institute of Mathematical Statistics},
  author = {Xiao,  Luo},
  year = {2019},
  month = Jan 
}

@article{Wooldridge1999,
  title = {{Asymptotic Properties of Weighted M-estimators for variable probability samples}},
  volume = {67},
  ISSN = {1468-0262},
  url = {http://dx.doi.org/10.1111/1468-0262.00083},
  DOI = {10.1111/1468-0262.00083},
  number = {6},
  journal = {Econometrica},
  publisher = {The Econometric Society},
  author = {Wooldridge,  Jeffrey M.},
  year = {1999},
  month = Nov,
  pages = {1385–1406}
}

@misc{NHANES,
	title = {{National Health and Nutrition Examination Survey : estimation procedures, 2011–2014}},
	url = {https://stacks.cdc.gov/view/cdc/51180},
	shorttitle = {National Health and Nutrition Examination Survey},
	author = {Chen, Te-Ching and Parker, Jennifer D. and Clark, Jason and Shin, Hee-Choon and Rammon, Jennifer R. and Burt, Vicki L.},
        year={2025},
	urldate = {2025-01-19},
	langid = {english}
}

@article{NHATS,
	title = {Cohort Profile: The National Health and Aging Trends Study ({NHATS})},
	volume = {48},
	issn = {0300-5771},
	url = {https://doi.org/10.1093/ije/dyz109},
	doi = {10.1093/ije/dyz109},
	shorttitle = {Cohort Profile},
	pages = {1044--1045g},
	number = {4},
	journal = {International Journal of Epidemiology},
	shortjournal = {International Journal of Epidemiology},
	author = {Freedman, Vicki A and Kasper, Judith D},
	urldate = {2025-01-19},
	date = {2019-08-01},
        year={2025},
	}

@book{fda_book,
  title={{Functional Data Analysis with R}},
  author={Crainiceanu, C.M. and Goldsmith, J. and Leroux, A. and Cui, E.},
  year={2024},
  publisher={Springer New York, NY, USA}
}

@book{Korn1999,
  title = {{Analysis of Health Surveys}},
  ISBN = {9781118032619},
  url = {http://dx.doi.org/10.1002/9781118032619},
  DOI = {10.1002/9781118032619},
  publisher = {Wiley},
  author = {Korn,  Edward L. and Graubard,  Barry I.},
  year = {1999},
  month = Jul 
}

@article{means2,
author = {Herv{\'e} Cardot and Camelia Goga and Pauline Lardin},
title = {{Uniform convergence and asymptotic confidence bands for model-assisted estimators of the mean of sampled functional data}},
volume = {7},
journal = {Electronic Journal of Statistics},
number = {none},
publisher = {Institute of Mathematical Statistics and Bernoulli Society},
pages = {562 -- 596},
year = {2013},
doi = {10.1214/13-EJS779},
URL = {https://doi.org/10.1214/13-EJS779}
}

@article{means3,
  title={Variance estimation and asymptotic confidence bands for the mean estimator of sampled functional data with high entropy unequal probability sampling designs},
  author={Cardot, Herv{\'e} and Goga, Camelia and Lardin, Pauline},
  journal={Scandinavian Journal of Statistics},
  volume={41},
  number={2},
  pages={516--534},
  year={2014},
  publisher={Wiley Online Library}
}

@article{medians,
  title = {{Using Complex Surveys to Estimate theL1‐Median of a Functional Variable: Application to Electricity Load Curves}},
  volume = {80},
  ISSN = {1751-5823},
  url = {http://dx.doi.org/10.1111/j.1751-5823.2011.00172.x},
  DOI = {10.1111/j.1751-5823.2011.00172.x},
  number = {1},
  journal = {International Statistical Review},
  publisher = {Wiley},
  author = {Chaouch,  Mohamed and Goga,  Camelia},
  year = {2012},
  month = apr,
  pages = {40–59}
}

@article{cui_fui,
	title = {Fast Univariate Inference for Longitudinal Functional Models},
	volume = {31},
	issn = {1061-8600},
	url = {https://www.ncbi.nlm.nih.gov/pmc/articles/PMC9197085
        /},
	doi = {10.1080/10618600.2021.1950006},
	pages = {219--230},
	number = {1},
	journal = {Journal of computational and graphical statistics : a joint publication of American Statistical Association, Institute of Mathematical Statistics, Interface Foundation of North America},
	shortjournal = {J Comput Graph Stat},
	author = {Cui, Erjia and Leroux, Andrew and Smirnova, Ekaterina and Crainiceanu, Ciprian M.},
	urldate = {2024-09-27},
	year = {2022},
	pmid = {35712524},
	pmcid = {PMC9197085},
}

@article{Crainiceanu_2012,
	title = {Bootstrap-based inference on the difference in the means of two correlated functional processes},
	volume = {31},
	pages = {3223-3240},
	number = {26},
	journal = {Statistics in Medicine},
	author = {Crainiceanu, C.M. and Staicu, A.M. and Ray, S. and Punjabi, N.},
	year = {2012}
}

@article{Morris_Carroll_2006,
	title = {Wavelet-based functional mixed models},
	volume = {68},
	pages = {179-199},
	number = {2},
	journal = {Journal of the Royal Statistical Society Series B},
	author = {Morris, J.S. and Carroll, R.J.},
	year = {2006}
}

@article{Morris_2008,
    author = {Morris, Jeffrey S. and Brown, Philip J. and Herrick, Richard C. and Baggerly, Keith A. and Coombes, Kevin R.},
    title = {Bayesian Analysis of Mass Spectrometry Proteomic Data Using Wavelet-Based Functional Mixed Models},
    journal = {Biometrics},
    volume = {64},
    number = {2},
    pages = {479-489},
    year = {2008},
}

@article{Park_Staicu_2018,
	title = {Simple fixed-effects inference for complex functional models},
	volume = {19},
	pages = {137-152},
	number = {2},
	journal = {Biostatistics},
	author = {Park, S.Y. and Staicu, A.-M. and Xiao, L. and Crainiceanu, C.M.},
	year = {2018}
}

@article{survey_sofr,
	title = {Scalar-on-function regression: Estimation and inference under complex survey designs},
	volume = {43},
	issn = {1097-0258},
	doi = {10.1002/sim.10194},
	shorttitle = {Scalar-on-function regression},
	pages = {4559--4574},
	number = {23},
	journal = {Statistics in Medicine},
	shortjournal = {Stat Med},
	author = {Smirnova, Ekaterina and Cui, Erjia and Tabacu, Lucia and Leroux, Andrew},
	year = {2024},
	pmid = {39136337},
}

@article{parker_bayesian_2023,
	title = {A Bayesian Functional Data Model for Surveys Collected under Informative Sampling with Application to Mortality Estimation Using {NHANES}},
	volume = {79},
	issn = {0006-341X},
	url = {https://doi.org/10.1111/biom.13696},
	doi = {10.1111/biom.13696},
	pages = {1397--1408},
	number = {2},
	journal = {Biometrics},
	shortjournal = {Biometrics},
	author = {Parker, Paul A. and Holan, Scott H.},
	urldate = {2024-10-15},
	year = {2023},
}

@article{lumley_software,
  title = {{Analysis of Complex Survey Samples}},
  volume = {9},
  ISSN = {1548-7660},
  url = {http://dx.doi.org/10.18637/jss.v009.i08},
  DOI = {10.18637/jss.v009.i08},
  number = {8},
  journal = {Journal of Statistical Software},
  publisher = {Foundation for Open Access Statistic},
  author = {Lumley,  Thomas},
  year = {2004}
}

@article{Lumley2017,
  title = {{Fitting Regression Models to Survey Data}},
  volume = {32},
  ISSN = {0883-4237},
  url = {http://dx.doi.org/10.1214/16-STS605},
  DOI = {10.1214/16-sts605},
  number = {2},
  journal = {Statistical Science},
  publisher = {Institute of Mathematical Statistics},
  author = {Lumley,  Thomas and Scott,  Alastair},
  year = {2017},
  month = may 
}

@Inbook{ppswor,
author="Singh, Sarjinder",
title="Use of Auxiliary Information: Probability Proportional to Size and Without Replacement (PPSWOR) Sampling",
bookTitle="Advanced Sampling Theory with Applications: How Michael ` selected' Amy Volume I",
year="2003",
publisher="Springer Netherlands",
address="Dordrecht",
pages="349--528",
isbn="978-94-007-0789-4",
doi="10.1007/978-94-007-0789-4_5",
url="https://doi.org/10.1007/978-94-007-0789-4_5"
}

@book{wahba1990spline,
  title={Spline models for observational data},
  author={Wahba, Grace},
  year={1990},
  publisher={SIAM}
}

@article{eilers1996flexible,
  title={{Flexible smoothing with B-splines and penalties}},
  author={Eilers, Paul HC and Marx, Brian D},
  journal={Statistical science},
  volume={11},
  number={2},
  pages={89--121},
  year={1996},
  publisher={Institute of Mathematical Statistics}
}

@article{osullivan,
  title={Nonparametric estimation of relative risk using splines and cross-validation},
  author={O’Sullivan, Finbarr},
  journal={SIAM Journal on Scientific and Statistical Computing},
  volume={9},
  number={3},
  pages={531--542},
  year={1988},
  publisher={SIAM}
}

@book{Ruppert2003,
  title = {{Semiparametric Regression}},
  ISBN = {9780511755453},
  url = {http://dx.doi.org/10.1017/CBO9780511755453},
  DOI = {10.1017/cbo9780511755453},
  publisher = {Cambridge University Press},
  author = {Ruppert,  David and Wand,  M. P. and Carroll,  R. J.},
  year = {2003},
  month = jul 
}

@book{Wood2017,
  title = {{Generalized Additive Models: An Introduction with R}},
  ISBN = {9781315370279},
  url = {http://dx.doi.org/10.1201/9781315370279},
  DOI = {10.1201/9781315370279},
  publisher = {Chapman and Hall/CRC},
  author = {Wood,  Simon N.},
  year = {2017},
  month = may 
}

@article{survey_reg,
  title = {A design-sensitive approach to fitting regression models with complex survey data},
  volume = {12},
  ISSN = {1935-7516},
  url = {http://dx.doi.org/10.1214/17-SS118},
  DOI = {10.1214/17-ss118},
  number = {none},
  journal = {Statistics Surveys},
  publisher = {Institute of Mathematical Statistics},
  author = {Kott,  Phillip S.},
  year = {2018},
  month = jan 
}

@article{BRR,
 ISSN = {00063444, 14643510},
 URL = {http://www.jstor.org/stable/2673521},
 author = {J.N.K. Rao and J. Shao},
 journal = {Biometrika},
 number = {2},
 pages = {403--415},
 publisher = {[Oxford University Press, Biometrika Trust]},
 title = {{Modified Balanced Repeated Replication for Complex Survey Data}},
 urldate = {2025-05-12},
 volume = {86},
 year = {1999}
}

@article{RWYB,
AUTHOR = {Beaumont, Jean-Francois and Emond, Nelson},
TITLE = {{A Bootstrap Variance Estimation Method for Multistage Sampling and Two-Phase Sampling When Poisson Sampling Is Used at the Second Phase}},
JOURNAL = {Stats},
VOLUME = {5},
YEAR = {2022},
NUMBER = {2},
PAGES = {339--357},
URL = {https://www.mdpi.com/2571-905X/5/2/19},
ISSN = {2571-905X},
DOI = {10.3390/stats5020019}
}

@Misc{paxminh,
  AUTHOR = {{Centers for Disease Control and Prevention}},
  HOWPUBLISHED = {\url{https://wwwn.cdc.gov/Nchs/Nhanes/2013-2014/PAXMIN_H.htm}},
  TITLE = {{NHANES} 2013-2014: Physical {Activity} {Monitor} {Data} {(PAXMIN\_H)}},
  YEAR = {2013},
note      = {Accessed: 2024-05-01}
}

@Misc{paxming,
  AUTHOR = {{Centers for Disease Control and Prevention}},
  HOWPUBLISHED = {\url{https://wwwn.cdc.gov/Nchs/Nhanes/2011-2012/PAXMIN_G.htm}},
  TITLE = {{NHANES} 2011-2012: Physical {Activity} {Monitor} {Data} {(PAXMIN\_G)}},
  YEAR = {2011},
    note    = {Accessed: 2024-05-01}
}

@article{mims,
  title = {{An Open-Source Monitor-Independent Movement Summary for Accelerometer Data Processing}},
  volume = {2},
  ISSN = {2575-6613},
  url = {http://dx.doi.org/10.1123/jmpb.2018-0068},
  DOI = {10.1123/jmpb.2018-0068},
  number = {4},
  journal = {Journal for the Measurement of Physical Behaviour},
  publisher = {Human Kinetics},
  author = {John,  Dinesh and Tang,  Qu and Albinali,  Fahd and Intille,  Stephen},
  year = {2019},
  month = dec,
  pages = {268–281}
}

@article{survey_mixed,
	title = {Linear mixed models for complex survey data: Implementing and evaluating pairwise likelihood},
	volume = {13},
	rights = {© 2024 The Authors. Stat published by John Wiley \& Sons Ltd.},
	issn = {2049-1573},
	url = {https://onlinelibrary.wiley.com/doi/abs/10.1002/sta4.657},
	doi = {10.1002/sta4.657},
	shorttitle = {Linear mixed models for complex survey data},
	pages = {e657},
	number = {1},
	journal = {Stat},
	author = {Lumley, Thomas and Huang, Xudong},
	urldate = {2025-01-19},
	year = {2024},
	langid = {english},
	note = {\_eprint: https://onlinelibrary.wiley.com/doi/pdf/10.1002/sta4.657},
}

@manual{SAS2023,
  author       = {{SAS Institute Inc.}},
  title        = {The SURVEYREG Procedure},
  year         = {2023},
  note         = {In \textit{SAS/STAT\textsuperscript{\textregistered} 15.3 User’s Guide}},
  address      = {Cary, NC},
  organization = {SAS Institute Inc.},
  url          = {https://documentation.sas.com/doc/en/statug/15.2/statug_surveyreg_toc.htm}
}

@manual{Stata2025,
  author       = {{StataCorp}},
  title        = {Stata 19 Survey Data Reference Manual},
  year         = {2025},
  address      = {College Station, TX},
  publisher    = {Stata Press},
  url          = {https://www.stata.com/manuals/svy.pdf}
}

@Misc{svrep,
    author = {Benjamin Schneider},
    year = {2023},
    title = {svrep: Tools for Creating, Updating, and Analyzing Survey Replicate Weights},
    note = {R package version 0.6.0},
    url = {https://CRAN.R-project.org/package=svrep},
  }

@article{mfpca,
  title = {Multilevel functional principal component analysis},
  volume = {3},
  ISSN = {1932-6157},
  url = {http://dx.doi.org/10.1214/08-AOAS206},
  DOI = {10.1214/08-aoas206},
  number = {1},
  pages = {458–488},
  journal = {The Annals of Applied Statistics},
  publisher = {Institute of Mathematical Statistics},
  author = {Di,  Chong-Zhi and Crainiceanu,  Ciprian M. and Caffo,  Brian S. and Punjabi,  Naresh M.},
  year = {2009},
  month = mar 
}

@article{lfpca,
  title = {Longitudinal functional principal component analysis},
  volume = {4},
  journal = {Electronic Journal of Statistics},
  author = {Greven, S. and Crainiceanu, C.M. and Caffo, B. and Reich, D.},
  pages = {1022-1054},
  year = {2010}
}

@article{sfpca,
  title = {Structured functional principal component analysis},
  volume = {71},
  number = {1},
  journal = {Biometrics},
  author = {Shou, H. and Zipunnikov, V. and Crainiceanu, C.M. and Greven S.},
  pages = {247-257},
  year = {2015}
}



\section*{Supporting Information}
Web Appendix A: FUI for non-longitudinal data \\
Web Appendix B: R code and detailed description of simulation \\
Web Appendix C: Analytical simulation vignette \\
Web Appendix D: Supplemental Tables and Figures \\
Web Appendix E: Sensitivity Analyses \\
\vspace*{-8pt}

\end{document}